\documentclass{jfm}
\usepackage[mylang]{appendix}
\usepackage{jfmcaps,epsfig,psfig,natbib}
\usepackage[english]{babel}
 
\def\dynpercm{\nobreak\mbox{$\;$dynes\,cm$^{-1}$}}

\def\cmpers2{\nobreak\mbox{$\;$cm\,s$^{-2}$}}
\def\gmperc3{\nobreak\mbox{$\;$gm\,cm$^{-3}$}}
\def\mpers{\nobreak\mbox{$\;$m\,s$^{-1}$}}

\def\Pec{\mbox{\rm Pe}}   % Peclet number
\newcommand{\beq}{\begin{equation}}
\newcommand{\eeq}{\end{equation}}
                                         
\ifCUPmtlplainloaded
\else
\fi
\ifCUPmtlplainloaded

  \renewcommand{\simeq}{\approx}
\fi
\ifCUPmtlplainloaded
  \font\bit = mtmib10 at 10.5pt \skewchar\bit ='177  % bold math italic
\else
  \font\bit = cmmib10 \skewchar\bit ='177  % bold math italic
\fi
\ifCUPmtlplainloaded
\else
  \font\tenbmi=cmmib10 at 10pt  \skewchar\tenbmi ='177
  \font\sevenbmi=cmmib10 at 7pt \skewchar\sevenbmi ='177
  \font\fivebmi=cmmib10 at 5pt  \skewchar\fivebmi ='177
 
  \newfam\bmifam
  \textfont\bmifam=\tenbmi
  \scriptfont\bmifam=\sevenbmi
  \scriptscriptfont\bmifam=\fivebmi
  
\fi
\newsavebox{\thalfbox}
\sbox{\thalfbox}{$\textstyle\frac{1}{2}$}

\newsavebox{\shalfbox}
\sbox{\shalfbox}{$\scriptstyle\frac{1}{2}$}

\newsavebox{\squartbox}
\sbox{\squartbox}{$\frac{1}{4}$} %RM removed scriptstyle

\newsavebox{\etbox}
\sbox{\etbox}{\boldmath$\eta$}
\newsavebox{\astrutbox}
\sbox{\astrutbox}{\rule[-5pt]{0pt}{20pt}}

\mathchardef\varLambda="0103
 
\ifCUPmtlplainloaded
\else
\fi
\title[Wind-driven gravity waves]
{Weakly non-linear analysis of wind-driven gravity waves}

\author[Alexakis, Young and Rosner]
{Alexandros Alexakis$^1$, Yuan-Nan Young $^2$ and Robert Rosner$^3$} 

\affiliation{$^1$ Department of Physics, University of Chicago, 
             Chicago, IL 60637, USA \\
             $^2$ Department of Engineering Sciences \&  Applied Mathematics,
             Northwestern University, Evanston, IL, 60208, USA\\
             $^3$ Departments of Physics and of Astronomy \& Astrophysics, 
             University of Chicago, Chicago, IL 60637, USA\\[\affilskip]}

\begin{document}

\maketitle

\begin{abstract}

We study the weakly non-linear development of shear-driven gravity waves,
following the physical mechanism first proposed by Miles and furthermore
investigate the mixing properties of the finite amplitude solutions. 
Calculations to date have been restricted to the linear theory, which predicts that
gravity waves are amplified by an influx of energy through the critical
layer, where the velocity of the wind equals the wave phase velocity.
Because of the presence of a critical layer, ordinary weakly non-linear
methods fail; in this paper, we use a rescaling at the critical layer and
matched asymptotics to derive an amplitude equation for the most unstable
wave, under the simplifying assumption that the physical domain is periodic.
These amplitude equations are solved numerically, in their
quasi-steady limit, for the cases of small density ratio (applicable to
oceanography), and for arbitrary density ratio but strong stratification
(for more general physical/astrophysical situations).
As is found in other analysis for critical layers in inviscid
parallel flow, we find that the
initial exponential increase of the amplitude $A$ transitions to an
algebraic growth rate proportional to the viscosity, $A \sim \nu t^{2/3}$.
However, for the air over water case (provided the maximum wind velocity is
in the range of $0.2\mpers \sim 1\mpers$),
our results from the weakly non-linear analysis for
the single mode show that the transition from exponential to
algebraic growth rate occurs when the amplitude of the wave
is as small as $h\sim 10^{-5}\lambda$; hence,
it may be difficult to observe the linear regime for this case in numerical
simulations.  (Whether this result transfers to cases in which the physical
domain is not periodic, so that the marginally stable modes form a continuum,
is not as yet known.)  We also find that the weakly non-linear
flow allows for super-diffusive particle transport with an 
exponent $\sim 3/2$, consistent with Venkataramani's results.

\end{abstract}

\pagebreak

\section{Introduction}

The generation of surface waves by winds has been a problem under study
for well over a century. The simple model of Kelvin-Helmholtz (from now
on, KH) instability provided higher bounds on the maximum wind
velocity for the instability to occur than what was experimentally
observed (\cite{Chandra62}, \S101); and a solution was not available until
\cite{Miles57,Miles59a,Miles59b,Miles62,Miles67} proposed a model
in which gravity ocean waves are amplified through a `resonant'
interaction with the wind above the ocean surface: Waves that travel with
speed $c=\sqrt{g/k}$ are amplified through an influx of energy at the
height $y_c$ in the wind where the wind speed $U(y)$ matches the wave
speed, $U(y=y_c)=c$. Miles explicitly assumed that the wind profile above
the sea surface was of the form $\log (z/z^*)$ (where $z^*$ is a
parameter), composed of a mean velocity $U$ and a turbulent component
$u$. Assuming the mean wind profile to be given from turbulent boundary
layer theory and the turbulent component to be small, he calculated the
influx of energy to the wave. This analysis involved the treatment of the
critical layer (the point where the velocity of the wind was equal to the
velocity of the gravity wave), where the solution of the linear
eigenfunction problem becomes singular (the $x$-component of the
perturbation velocity behaves as $u \sim \log (y-y_c)$). The singularity
is removed either because there is an imaginary component of $c$ (i.e.,
unsteady critical layer) or because viscosity becomes important in this
thin layer (i.e., viscous critical layer). In both cases, the final
result is that there is an `$-i\pi$ phase change' across the critical
layer, e.g., the perturbation wave above the critical layer is not in
phase with
the wave below. A direct result of this phase change is that the gravity
wave will become unstable because a component of the
pressure perturbation will be in phase with the slope of the wave
(unlike the KH case, in which the pressure perturbation wave is always in
phase with the wave). Using the results  from linear theory and for small
density ratios, Miles predicted the energy flux from the wind to the
gravity wave.

Our interest in this problem is motivated by an astrophysical puzzle,
namely, the mixing of carbon/oxygen (C/O) at the surface of a white
dwarf with accreted material (mostly hydrogen and helium, He/H)
overlying the stellar surface (whose presumed origin is from an
accretion disk surrounding the compact star). For a variety of reasons,
it is thought that the accreted envelope is in differential rotation with
respect to the stellar surface, so that a `wind' is expected at this
surface \citep{Rosner01}. In this stellar case, one is forced to generalize
the earlier results to arbitrary density ratios (between the
`atmosphere' and the `surface material'); and we have already done so
for the linear problem \citep{Alexakis01}, deriving bounds on the
instability in the parameter space, and estimating the growth rates of
the unstable modes. While useful as a starting point for further
analysis, linear theory gives no information about the ultimate fate of
the system, which is largely governed by nonlinear processes. 
In this paper, we uncover the non-linear effects
on the evolution of wind-driven surface waves by examining 
their finite-amplitude evolution.  To explore the applications
to the relevant astrophysical and geophysical systems, 
we also investigate the mixing property of the weakly 
non-linear flow using particle method,
and results show enhanced mixing due to the coupling of the 
weakly nonlinear wind with the surface wave.

We note here that the weakly non-linear theory for the KH instability has
already been derived by \cite{Drazin70}. That analysis cannot be
applied straightforwardly to our problem because of the presence of
critical layers: Due to the singularity that appear in the linear theory
at the point where the phase speed of a surface wave matches the wind
speed, higher order terms in the expansion become more singular and the
expansion must ultimately fail. The fundamental reason for this behavior
is that the flow becomes nonlinear first inside the critical layer even
though the rest of the flow can still be considered as operating in
the linear regime. For this reason, a more refined treatment of the
critical layer is required. The necessary analytical `machinery'
fortunately already exists: Thus, \cite{Benney69}, and
later \cite{Benney75}, observed that for small but
finite amplitudes the phase change at the critical layer does not
necessarily have to be $-i\pi$; instead, they found a solution with the
property that if the nonlinear terms are taken into account, the phase
change is zero. Later \cite{Haberman72} showed
numerically that there is a smooth increase of the phase change, from
$-i\pi$ to 0, and introduced the function $\Phi$ that gives the phase
change as a function of the amplitude; 
\cite{Churilov87,Churilov88,Churilov96} then developed a weakly nonlinear
theory based on $\Phi$ and other similar functions defined for the
appropriate critical layer problem. A fundamental assumption in all this
work is that the viscosity is dominant in the critical layer; this leads
to the derivation of an ordinary differential equation for the wave
amplitude. The full equations of the weakly nonlinear problem without
making the previous assumption have then been solved numerically for
various cases \citep{Goldstein88,Warn78,Balmforth01}.

In our case the treatment -- although closely related to the earlier work
-- is nevertheless different in two respects. The first one is that the
previous cases possess stability boundaries for modes for which the
critical layer is formed at the inflection point $\partial^2_{y}U/
\partial y^2 =0$; one then examines the close-to-marginally-stable state
unstable
modes. This is however not our case since such an inflection point does
not exist (unless one considers $U(\infty)$ as such a point). Instead, we
will be considering two special cases: the case of small density ratio,
and the case of very strong stratification\footnote{This case is of
particular interest to our astrophysical problem, since the instability
takes place on the surface of a white dwarf star, i.e., a star of solar
mass, but comparable in size to the Earth.}; in both cases, the linear
growth rate is small. This procedure will lead to a slightly different
scaling.
The second difference lies in the fact that we have an interface. Because
of this, our solvability condition will not be expressed in terms of
integrals but rather in terms of  appropriate vector products of the
values of the perturbation stream function at the interface.

This paper is structured as follows. First we formulate our problem and
describe the non-dimensionalized form of the basic problem equations.
In \S 3 we very briefly review (and slightly revise) the linear theory
developed earlier \citep{Alexakis01}. The amplitude equations are derived
in \S 4.1 and \S4.2 for the small density ratio and the strongly
stratified cases, respectively (we define these terms more precisely in
what follows). We summarize the conservation laws from
the amplitude equations in \S \ref{prelim}, and we discuss the implications of the
long-time behavior of surface waves in \S \ref{quasi}.  
In \S \ref{nume} we summarize results from numerically simulating 
the amplitude equations.  In \S 6 we investigate how the 
coupling between the wind and the surface wave affects mixing 
properties of the finite amplitude solution.  
A detailed examination of the assumptions made in the analysis is provided
in \S 7, where we also draw conclusions from our work.

\section{Formulation}

We consider a two-layer system with constant fluid density $\rho_1$ and
$\rho_2$ ($\rho_1 \le \rho_2$), respectively, in the upper and lower layers.
The interface between the two layers is given by $y=h(x,t)$, where $x$
is the horizontal and $y$ is the vertical coordinate. In the upper
layer we assume a wind parallel to the originally flat interface $y=h(x,t=0)=0$.
The wind has a shear velocity profile $\vec U \equiv [U(y), 0]$,
where $U(y)$ varies only with $y$.
The lower layer is initially at rest.
Our aim is to study the dynamics and the weakly
nonlinear development of a small sinusoidal perturbations of the horizontal
interface.\footnote{As an aside, we note that the presence of a sharp
interface boundary between the two fluids in our system is really
only a device to simplify the analysis, but in no way restricts our
results. That is, in a more general case there will be no sharp
interface, but the density will change smoothly from $\rho_2$ to $\rho_1$
in a layer of width $\delta_1$. We would then also expect $U(y)$ to be a
smooth function in $y$, so that any flow discontinuity at the interface
would be replaced by a smooth variation in $U$ within a thin viscous
boundary layer of width $\delta_2$. One would then expect to see
differences in behavior only for those modes with horizontal wavenumber
$k$ large enough so that the critical layer lies inside these layers
$\delta_1,\delta_2$. Such modes, however, are KH-modes which are not
under study here; they will obey a Richardson-type stability criterion,
$1/4 < g (\Delta \rho/\rho)\delta_2^2/\delta_1$ \citep{Chandra62}. We
therefore expect our assumption of a sharp interface to be reasonable for
horizontal wave-numbers $k^{-1} \ge \max\{\delta_1,\delta_2\}$. Since we
are primarily interested in long wavelength perturbations, we conclude
that, for our purposes, we have not disregarded any important physics.}

The fluid is assumed incompressible in both layers; we therefore
work with the stream function $\Psi$, which is connected to the
velocity by the relation $(u,v)=(\partial_y \Psi, -\partial_x\Psi)$.
The stream function can be separated into
its mean and its perturbation components,

\begin{equation}
\Psi_{\mbox{{\scriptsize{total}}}}=\int_0^yUdy'+\Psi \,.
\end{equation}
The fluid within each layer is described by
the Navier-Stokes equations (in terms of the stream function $\Psi$):

\begin{equation}
\nabla^2\Psi_{,t}+U\nabla^2\Psi_{,x}-U_{,yy}\Psi_{,x}
=\Psi_{,x}\nabla^2\Psi_{,y}-\Psi_{,y}\nabla^2\Psi_{,x}+
\nu \nabla^2\nabla^2 \Psi ,
\label{NS}
\end{equation}
where $\nabla$ is the two-dimensional Laplacian
and $\nu$ is the viscosity (which we will consider to be small, and
therefore negligible except for a narrow region within the critical
layer). Here we have used the standard notational device of
comma-prefaced subscripts to denote partial derivatives, e.g.,
\[
f_{,x} \equiv \partial f / \partial x \,.
\]
The boundary conditions at the interface are the continuity
of the perpendicular component of the velocity to the interface

\begin{equation}
h_{,t}+(U^{\pm}+\Psi^{\pm}_{,y})h_{,x}+\Psi^{\pm}_{,x} = 0 \,,
\label{BC1}
\end{equation}
and the continuity of pressure
%\begin{equation}
\[ \Delta \Big{[}\rho_i\big{\{}
\Psi_{,ty}+(U+\Psi_{,y})\Psi_{,xy}
-\Psi_{,x}(U_{,y}+\Psi_{,yy})
-h_{,x}(\Psi_{,tx}+(U+\Psi_{,y})\Psi_{,xx}
+\Psi_{,x}\Psi_{,xy}) \big{\}}
\Big{]} \]
\begin{equation}
=gh_{,x}(\rho_2-\rho_1) \,,
\label{BC2}
\end{equation}
where $h(x)$ is the elevation of the interface and
all quantities above are evaluated at  $y=h(x)$. The $\pm$ indices
indicate values above and below the interface, and
$\Delta [ \quad ]$ denotes the difference across $y=h$
(e.g., $\Delta [f(y) ]= f(h^+)-f(h^-)$ ).

We non-dimensionalize lengths by the
characteristic length $l$ of the wind, and velocities by the
asymptotic value of the wind at $y \to +\infty$, $U_{max}$. An
important parameter that emerges from the scaling is $G \equiv gl/U^2$,
which is a measure of the ratio of potential energy to kinetic energy, or
alternatively, a measure of the strength of the stratification. Other
dimensionless parameters are the Reynolds number,
$Re \equiv Ul/\nu \gg 1$, and the ratio of densities,
$r \equiv \rho_1/\rho_2 \le 1$.
In terms of the non-dimensional parameters, we focus on cases
where $Re\gg 1$ in each layer, and especially the two distinguished
limits $r \ll 1$ (air on water, for example) and arbitrary `large'
density ratio $r$ with $G \gg 1$ (accretion on white dwarfs in
astrophysics).

\section{Linear Theory}

We linearize equations (\ref{NS}) - (\ref{BC2}), and write the stream
function perturbation $\Psi$ and surface elevation
$h$ as a traveling wave parallel to the wind
(right traveling wave in our setup)

\begin{eqnarray*}
\Psi^{\pm}(y,x,t) &=& \phi^{\pm}(y)e^{iK(x-Ct)}+ \mbox{c.c.} \,, \\
h &=&  \tilde{h}e^{iK(x-Ct)}+ \mbox{c.c.} \,,
\end{eqnarray*}
where $\mbox{c.c.}$ denotes the complex conjugate, and $K=kl$ is the
non-dimensional (horizontal) wavenumber and $C$ is the non-dimensional
phase velocity.
  This leads us to the well-studied
Rayleigh equation \citep{Drazin81}

\begin{equation}
\phi^{\pm}_{,yy}-\left[K^2+\frac{U_{,yy}}{U-C} \right]\phi^{\pm} = 0 \,,
\label{linear}
\end{equation}
where the inviscid limit $Re \to \infty$ is taken.
The boundary conditions become
\begin{equation}
(U^{\pm}-C)\tilde{h}+\phi^{\pm}=0 \,,
\label{BCL1}
\end{equation}
\begin{equation}
\Delta \big{[} \rho_i( (U^{\pm}-C)\phi_{,y} -U^{\pm}_{,y} \phi^{\pm})
\big{]}
-G\tilde{h}(\rho_2-\rho_1)=0 \,,
\label{BCL2}
\end{equation}
Eliminating $\phi^-$ and $\tilde{h}$ from the last two equations, and
dropping the + index for convenience, we obtain
\begin{equation} 
KC^2-r\Big{[}(U-C)^2 \phi_{,y}-(U-C)U_{,y}
\Big{]}-G(1-r)=0 \,,
\label{BCL12}
\end{equation}
where we have used $\phi^+|_{y=0}=1$ as a normalization condition.

To further simplify we assume that $U(0)=0$ and omit
the KH-modes (of which the nonlinear evolution is not
under study here, see \cite{Alexakis01} for more details).
The linear growth rate ($K\mbox{Im}(C)$) has been numerically calculated
and summarized in \cite{Alexakis01} for cases of interests.
Here we briefly summarize some relevant results on stability
boundaries for the wind profile

\begin{equation}
\label{exp_wind}
U=1-e^{-y} \,.
\end{equation}
As shown in \cite{Alexakis01}, this wind profile allows for
an analytic expression for the stability boundaries in the
stability diagram.  One can show that some general features
of the stability boundaries summarized here for wind profile in
equation (\ref{exp_wind}) also hold for other bounded wind profiles.
The stability bound comes from the modes that
have phase velocity $C=1$, in which case the solution of
equation (\ref{linear}) becomes
$\phi=e^{i\kappa x}$ with $\kappa=\sqrt{1+K^2}$. Applying the boundary
conditions, one obtains the criterion
on wavenumber $K$ for the unstable modes:

\begin{equation}
K \ge K_{min}=
\frac{G(1-r)+r-r\sqrt{(G(1-r)+r)^2+(1-r^2)}}{1-r^2} \quad .
\label{BOUND}
\end{equation}

\noindent
The unstable modes can be further restricted if we assume the presence of
surface tension or magnetic fields. This will lead to an additional term
${\mathcal T}K^2$ in equation (\ref{BCL12}), where
${\mathcal T}=\sigma/(\rho_2U_{max}^2l)$ and $\sigma$ is the surface
tension\footnote{We note that a magnetic field whose direction is aligned
with the interface will have the same effect as surface tension with
$\sigma(K)=B^2/(2\pi \mu K)$ \citep{Chandra62}.}. Then the
unstable modes lie in the region $K_{min} \le K \le K_{max}$, where
$K_{min}$ and $K_{max}$ are given by the positive real solutions of

\begin{equation}
K+r\sqrt{1+K^2}-\big{(}G(1-r)+r+{\mathcal T}(K)K^2\big{)}=0 \,.
\label{BOUNDT}
\end{equation}
Furthermore, there is a minimum value of ${\mathcal T}$ below which the
above equation has no positive real solutions, and therefore no unstable modes
exist. The physics behind these bounds is simple: In order for a mode to
become unstable for the above wind profile, the phase velocity of the wave
must lie in the range $0<c<U_{max}$ \citep{Alexakis01}. With the inclusion
of surface tension, the phase velocity is not monotonic with $K$, but for
large enough $K$, the phase velocity increases in an unbounded fashion.
This leaves only a finite  region in $K$ space with phase velocity smaller
than $U_{max}$; moreover, if the surface tension is large enough, the
minimum phase velocity is larger than $U_{max}$, and therefore no unstable
mode exists.

The analysis is simplified if we assume small density ratio, and expand
all quantities in $r$:
\[ C=C_0+rC_1+\dots\]
\[ \phi=\phi_0+r\phi_1+\dots\]
In that case the linear theory, to zero order, gives a gravity wave with
phase velocity $C_0=\sqrt{G/K}$. At the next order, one obtains
\begin{equation}
2C_0C_1=[C_0^2\phi_{0,y}+U_{,y}]-G \,,
\end{equation}
and therefore
\begin{equation}
\mbox{Im}\{C_1\}=\frac{1}{2}C\mbox{Im}\{\phi_{0,y}\} \,
\end{equation}
where $\phi_0$ is such that
\begin{equation}
\phi_{0,yy}-\left[K^2+\frac{U_{,yy}}{U-C_0}\right]\phi_0 = 0 \,.
\end{equation}
An $-i\pi$ phase change at the critical level is assumed. Due to this
phase change, $\phi_{0,y}$ is complex at the
interface and thus $C_1$ has an imaginary component, which is
responsible for the instability of the traveling waves.

For the arbitrary $r$ case, the previous expansion does not hold, and
one needs to solve the full set of equations (\ref{linear})-(\ref{BCL12})
as in \cite{Alexakis01}. There it was shown that the
growth rate exhibits an exponential dependence on the parameter $G$. In
Appendix A, we carry out an asymptotic analysis for large $G$, and
derive this exponential dependence. More specifically, the growth
rate is found to be

\begin{equation}
K ~ \mbox{Im}\left\{C\right\}
=\frac{r\pi^2}{4(1-r)}\frac{1}{{\mathcal{A}}_tG}
\left(\frac{{\mathcal{A}}_tG}{K}\right)^{3/2}
\left[\left(1-\frac{1}{\sqrt{K/({\mathcal{A}}_tG)}}\right)^
{(K/{\mathcal{A}}_tG)}
\right]^{2{\mathcal{A}}_tG} \,,
\label{exp}
\end{equation}
where ${\mathcal{A}}_t$ is the Atwood number. As is shown in the
Appendix, the stream function $\Psi$ below the critical layer is composed
of an exponentially increasing and an exponentially decreasing component.
Since the boundary condition (equation (\ref{BCL1})) must be satisfied, the
exponentially large component must be in phase with $h$, which
leaves us with the exponentially small, out of phase, component to
drive the wave unstable. A detailed calculation then leads to the result
given above.

%%%%%%%%%%%%%%%%%%%%%%%%%%%%%%%%%%%%%%%%%%%%%%%%%%%%%%%%%%%%%%%%

\section{Weakly nonlinear theory}

We are now ready to embark upon the weakly non-linear theory. Formally
this is done by assuming that, for some parameter ranges of interests,
our physical system lies close to a marginally stable state so that an 
asymptotic expansion is allowed near the center manifold. 
For the problem at hand, however, and in the absence
of surface tension (${\mathcal T}=0$), marginally stable states are possible only when $r=0$
or $1/G=0$  (we note that $K$ is not a control parameter); the first one
expresses the unphysical situation that there is no upper fluid, and
the latter corresponds to a situation where there is no wind in the upper
fluid ($U\sim 1/\sqrt{G}$). Complication arises as we
deviate from these neutrally stable states.  In ordinary
dissipative systems, only a small number of modes near the center
manifold become unstable and need to be considered. 
In our case though, once the density ratio $r$ or the parameter $G$ is
finite, an infinite number of modes become unstable 
if surface tension ${\mathcal T}=0$.
Ideally the interaction of all these modes needs to be taken into
account. Practically this difficulty is removed by the combined effect
of surface tension ${\mathcal T}$ (or, equivalently, the presence of a
magnetic field) and weak viscous damping.
In the presence of surface tension the stability boundary
in the $(r,G,{\mathcal T})$ space is given
by the condition for positive solutions
of eq.(\ref{BOUNDT}) for $r=0$. Surface tension reduces the number of
unstable modes by neutralizing modes of wave numbers above some cut-off value.
Further more weak viscosity will damp out the neutrally stable modes of high wave numbers, 
rendering them asymptotically stable.  Using a periodic domain we can always fix the
period so that only one mode becomes unstable. We can then derive the
amplitude equation for this single mode as usual.

We find that these amplitude equations are almost the same
as those obtained from asymptotic expansion around the most unstable mode
without surface tension; the only difference is a coefficient
dependence on the surface tension. In the following derivation, for
simplicity, we chose not to include surface tension, and assume the
presence of surface tension only implicitly.
At the end of the analysis the surface tension can always be recovered by
replacing $G$ with $G+{\mathcal T}K^2$.
 
With the previously stated assumptions, and following the basic strategy
of the weakly nonlinear theory, we develop below an asymptotic expansion
based on the small amplitude of the perturbation, and introduce a
long time scale of the same order as the nonlinear terms inside the
critical layer. We consider two different cases: The first
case assumes that the density ratio is small ($r \ll 1$); the second case
allows for arbitrary $r$ but assumes that $G \gg 1$. 
The samll density ratio ($r$) in the first case
is the parameter to which we scale our amplitudes.
The large $G$ and arbitrary $r$ case, however, is a
bit more complicated; for this case, the growth rate according to linear
theory is proportional to $\exp(-2G)$, and this will be our scaling
parameter.

\subsection{Small density ratio  case}

We start with the small $r$ case. The following analysis applies to 
for a monotonicly increasing but otherwise arbitrary wind-profile.
%both logarithmic and exponential wind profiles in \cite{Alexakis01}. 
The outline of the derivation is as follows: we write the Euler equations,
and the boundary conditions, in the reference frame of the moving wave.
We introduce the scaling  discussed below, and solve up to the second
order; this requires a more detailed treatment at the critical layer that
determines the phase change across the layer. Using matched asymptotics,
we match the inner solution with the Frobenius solutions $\phi_a,\phi_b$
of the outer flow. This introduces five different amplitudes,
$A^{\pm}_1, B_1^{\pm}$, and $A_2$, for each Frobenius solution above and
below the critical layer (which will be assumed to be given numerically);
the fifth amplitude $A_2$ is just the surface wave amplitude. The first
four amplitudes are connected through the boundary conditions at infinity
and the matching at the critical layer; the fifth is connected with the
rest through the boundary conditions at the interface. When we apply the
boundary conditions, the first order gives the free gravity wave (vacuum
over water); at second order, the solvability condition results in the
amplitude equation.

We begin by introducing the scaling. Let $\epsilon \equiv \rho_1/\rho_2
\ll 1$ and $\partial_t \equiv \epsilon\partial_T-C\partial_x $. Then the
equation for the stream function in the reference frame of the wave will
be:

\begin{equation}
\epsilon\nabla^2\Psi_{,T}+(U-C)\nabla^2\Psi_{,x}-U_{,yy}\Psi_{,x}
=\Psi_{,x}\nabla^2\Psi_{,y}-\Psi_{,y}\nabla^2\Psi_{,x}+
\frac{1}{R}\nabla^2\nabla^2 \Psi \,,
\label{Navier}
\end{equation}
with boundary conditions at $y=0$

\begin{equation}
\epsilon h_{,T}-Ch_{,x}+\Psi^{\pm}_{,x} = \mbox{non-linear terms} \quad
\label{bound1}
\end{equation}
and
\begin{equation}
\epsilon\Big{[}  \epsilon\Psi_{,Ty}^+
-C\Psi^+_{,xy}-U^+_{,y}\Psi^+_{,x} \Big{]}
- \Big{[}\Psi_{T,y}^-C\Psi^-_{,xy}-U_{,y}\Psi^-_{,x} \Big{]}
-Gh_{,x}(1-\epsilon) =
\mbox{non-linear terms} \,,
\label{bound2}
\end{equation}
where the $\pm$ index means above or below the interface.
First we focus on the upper fluid and the outer solution
(outside the critical layer).

\subsubsection{Outer solution in upper fluid}

We expand the stream function as

\begin{equation}
\Psi=\epsilon^2\Psi_0+\epsilon^3\Psi_1 + \dots
\end{equation}
To zeroth order we have
\[
(U-C) \nabla^2\Psi_{0,x}-U_{,yy}\Psi_{0,x}=0 \,.
\]
Focusing on the most unstable mode and writing
$\Psi_0=A(T)\phi(y)e^{iKx}+ \mbox{c.c.}$, we obtain:
\[
\phi_{,y}-\left[K^2+\frac{U_{,yy}}{U-C} \right]\phi = 0 \,.
\]
Near the critical layer $y \to y_c$ the solution can be written as
\[
\phi=a^{\pm}\phi_a +b^{\pm}\phi_b \,.
\]
The $\pm$ signs correspond to above or below the critical layer, and
the two Frobenius solutions are
\begin{eqnarray}
\phi_a &=& \left(1+\frac{U_c''}{U_c'}(y-y_c)\ln|y-y_c|+... \right) \,, \\
\phi_b &=& \left(\frac{U_c''}{U_c'}(y-y_c)+... \right) \,,
\label{Frobenius}
\end{eqnarray}
where $U_c'$ and $U_c''$are $\partial U
/ \partial y |_{y=y_c}$ and $\partial^2 U / \partial y^2 |_{y=y_c}$,
respectively. In general, $a^{\pm}$ and $b^{\pm}$ are functions of $T$, so
that it is more convenient to write the solution as

\begin{equation}
\Psi_0=\big{[}A_{1\pm}(T)\phi_a(y)+B_{1\pm}(T) \phi_b(y)\big{]}e^{iKx} +
\mbox{c.c.}
\end{equation}
At the next order, we have
\[
(U-C)\nabla^2\Psi_{1,x}-U_{,yy}\Psi_1=-\nabla^2\Psi_{0,T} \,.
\]
Again expanding $\Psi_1$ in terms  of Frobenius solutions, we write $\Psi_1$ as
\[\Psi_1=\left(\hat{A}_{1\pm}\phi_a+\hat{B}_{1\pm}\phi_b+
A_{1\pm,T}\frac{U_c''}{U_c'^2} \ln |y|+...\right)e^{iKx}+\mbox{c.c.} \,,
\]
where $\hat{A}_{1\pm}$ and $\hat{B}_{1\pm}$ are the amplitudes of
$\Psi_1$. In order to match with the inner solution, we require the
behavior of $\Psi$ as $(y-y_c)\to \epsilon Y$. Upon expansion in $Y$ we
obtain

\begin{eqnarray*}
\Psi &=& \epsilon^2 \Psi_0 +\epsilon^3 \Psi_1+... \\
&=& \bigg{\{}
\epsilon^2A_{1\pm}
+\epsilon^3 \ln (\epsilon)\left[
A_{1\pm}\frac{U_c''}{U_c'}Y+\partial_TA_{1\pm}\frac{U_c''}{U_c'^2}
\right]+ \\
&\phantom{=}& ~~~~ +
\epsilon^3\left[
B_{1\pm}\frac{U_c''}{U_c'}Y+A_{1\pm}\frac{U_c''}{U_c'}Y \ln |Y|+
\partial_TA_{1\pm}\frac{U_{c}''}{U_c'^2} \ln |Y| \right]
+....    \bigg{\}}e^{iKx}+ \mbox{c.c.}
\end{eqnarray*}

%%%%%%%%%%%%%%%%%%%%%%%%%%%%%%%%%%%%%%%%%%%%%%%%%%%%%%%%%%%%%%%%%%%%%

\subsubsection{Inner solution}

To capture the dynamics inside the critical layer, we have to use the
scaling $\Psi  \to  \epsilon^2 \tilde{\Psi}(Y)$,
$y-y_c \to  \epsilon Y$ and $1/R   \to  \epsilon^3 \nu$.
  From equation (\ref{Navier}) we then obtain

%\begin{equation}
\[
\tilde{\Psi}_{,TYY}+U_c'Y\tilde{\Psi}_{,xYY}+
\tilde{\Psi}_{,Y}\tilde{\Psi}_{,YYx}-\tilde{\Psi}_{,x}\tilde{\Psi}_{,YYY}
-\nu \tilde{\Psi}_{,YYYY} \]
\begin{equation}
= -\epsilon \left[
\frac{1}{2}U_c''Y^2\tilde{\Psi}_{,YYx}- U_c''\tilde{\Psi}_{,x} \right]
+{\mathcal O}(\epsilon^2) \,.
\end{equation}
In order to match with the outer solution we expand $\tilde{\Psi}$ as
\[
\tilde{\Psi}=\tilde{\Psi}_0+\epsilon \ln (\epsilon)\tilde{\Psi}_1
+\epsilon\tilde{\Psi}_2+\epsilon^2 \ln (\epsilon)\tilde{\Psi}_3
+\epsilon^2\tilde{\Psi}_4 + ... \,.
\]
To first order, we then have
\[
\tilde{\Psi}_{0,TYY}+U_c'Y\tilde{\Psi}_{0,xYY}+
\tilde{\Psi}_{0,Y}
\tilde{\Psi}_{0,YYx}-\tilde{\Psi}_{0,x}\tilde{\Psi}_{0,YYY}
-\nu \tilde{\Psi}_{0,YYYY}=0 \,.
\]
Matching with the outer solution we obtain

\begin{equation}
\tilde{\Psi}_0=A_{1+}e^{iKx}+A_{1+}^*e^{-iKx} \,,
\label{Psi0}
\end{equation}
and therefore

\begin{equation}
A_{1+}=A_{1-}=A_{1} \,.
\label{A1}
\end{equation}
To second order $(\epsilon\ln(\epsilon))$, we have
\[
\tilde{\Psi}_{1,TYY}
+U_c'Y\tilde{\Psi}_{1,xYY}
-\tilde{\Psi}_{0,x}\tilde{\Psi}_{1,YYY}
-\nu \tilde{\Psi}_{1,YYYY}=0 \,.
\]
Matching with the outer solution we obtain
\begin{equation}
\tilde{\Psi}_1=\left[A_1\frac{U_c''}{U_c'}Y+\partial_TA_1
\frac{U_c''}{U_c'^2} \right]e^{iKx}+\mbox{c.c.}
\end{equation}
To third order ($\epsilon^3$), we have
\begin{equation}
\tilde{\Psi}_{2,YYT}+U_c'Y
\tilde{\Psi}_{2,YYx}-\tilde{\Psi}_{0,x}\tilde{\Psi}_{2,YYY}
-\nu \tilde{\Psi}_{2,YYYY}=U_c''\tilde{\Psi}_{0x} \,.
\end{equation}
Denoting $Z=\tilde{\Psi}_{2,YY}$, which is the vorticity inside the
critical layer, we obtain

\begin{equation}
Z_{,T}+U_c'YZ_{,x}-\tilde{\Psi}_{0,x}Z_{,Y}-\nu
Z_{,YY}=U_c''\tilde{\Psi}_{0,x} \,.
\label{Z}
\end{equation}
To match with the outer solution, we require the boundary conditions of $Z$

\[\lim_{Y\to\infty}Z=\left[ A_1\frac{U_c''}{U_c'}\frac{1}{Y}-
\frac{U_c''}{U_c'^2}\frac{1}{Y^2}A_{1T}\right]e^{iKx}+\mbox{c.c.} \]

\[\lim_{Y \pm\infty} \Psi_{2,Y} =
e^{iKx}\left[ A_1\frac{U''_c}{U'_c}( \ln |y|+1)
+A_{1T}\frac{U''_c}{U'_c}\frac{1}{Y}
+B_{1\pm}\frac{U''_c}{U'_c} \right]+ \mbox{c.c.} \]
Integrating $Z$ along $Y$, we obtain
\[
\int^{+\infty}_{-\infty}ZdY=\Big{[}\Psi_{2,Y}\Big{]}^{+\infty}_{-\infty}
=\frac{U''_c}{U'_c}[B_{1+}-B_{1-}]e^{iKx}+\mbox{c.c.} \,,
\]
where in the above integral we assumed the limiting procedure
$\lim_{\epsilon\to0}\int^{1/\epsilon}_{-1/\epsilon}Z ~dY$. Let

\begin{equation}
J=\frac{K}{2\pi}\frac{U'_c}{U''_c}\int^{+\pi/K}_{-\pi/K}e^{-iKx}
\int^{+\infty}_{-\infty}ZdYdx=B_{1+}-B_{1-} \,.
\label{J}
\end{equation}
This leads to the important result:

\begin{equation}
B_{1-}=B_{1+}-J \,.
\label{phaseB}
\end{equation}

The last equation (\ref{phaseB}) implies that the phase change across the
critical layer depends on the more detailed treatment of the vorticity
dynamics inside the critical layer. In the linear case $J=+i\pi A_1$, but
as the amplitude grows and the non-linear term $\tilde{\Psi}_{0,x}Z_{,Y}$
in equation (\ref{Z}) becomes important, the phase change is going to
decrease.

%%%%%%%%%%%%%%%%%%%%%%%%%%%%%%%%%%%%%%%%%%%%%%%%%%%%%%%%%%%%%%%%%%%%%%%

\subsubsection{Lower fluid}

Using the same scaling as for the outer solution of the upper fluid, we
have
\[\Psi^-=\epsilon^2\Psi_0^-+\epsilon^3\Psi_1^- + ...\]
To first order, we then have
\[
\nabla^2\Psi^-_x=0 \,,
\]
\[
\Psi^-=A_2(T) \tilde{\phi}_0^-
e^{iKx+Ky}+\mbox{c.c.}
\]
At next order,
\[
\nabla^2\Psi_{0,T}-C\nabla^2\Psi_{1,x}=0 \,,
\]
\[
\Psi_1^{-}=A_2'(T)\tilde{\phi}_1^-e^{iKx+Ky}+\mbox{c.c.}
\]
At this point we are ready to apply the boundary conditions at the
interface, and obtain the amplitude equation.

%%%%%%%%%%%%%%%%%%%%%%%%%%%%%%%%%%%%%%%%%%%%%%%%%%%%%%%%%%%%%%%

\subsubsection{Boundary conditions and the amplitude equation}

First, we expand the amplitude of the perturbed interface
\[
h=\epsilon^2h_0+\epsilon^3h_1+...
\]
To first order, we obtain
\[
h_0=A_2\tilde{h}_0e^{iKx}+\mbox{c.c.}
\]
Then using equations (\ref{bound1}) and (\ref{bound2}), we obtain

\begin{equation}
\tilde{\phi}_0^--C\tilde{h}_0=0 \,,
\label{b11}
\end{equation}

%and from (\ref{bound2})

\begin{equation}
-C\tilde{\phi}_0^-+G/K\tilde{h}_0=0 \,.
\label{b12}
\end{equation}
Let
\[
V_o=\left[
\begin{array}{c}
\tilde{h}_0 \\
\tilde{\phi}_0^-
\end{array}
\right] \quad \mbox{ and } \quad M=
\left[
\begin{array}{cc}
-C & 1 \\
    G/K & -C
\end{array}
\right]
\,.
\]
Then the previous set of equations (\ref{b11})-(\ref{b12}) can be written
as $MV_o=0$. For a non-trivial solution we must have det$[M]=0$; from
this condition we obtain

\begin{equation}
C=\sqrt{G/K} \,,
\end{equation}
and
\begin{equation}
\tilde{\phi}_0^-=C\tilde{h}_0 \,.
\end{equation}
  From equation (\ref{bound1}) we also obtain
\[
-Ch_{0,x}+\Psi_{0,x}^+|_0=0 \,,
\]
or

\begin{equation}
-C\tilde{h}_0A_2+(A_1\phi_a|_0+B_{1-}\phi_b|_0)=0 \,.
\label{b13}
\end{equation}
To the next order we have
\[ h_1=\hat{A}_2(T)\tilde{h}_1e^{iKx}+ \mbox{c.c.} \]
and
\begin{eqnarray*}
    \Psi_{1,x}^-|_0-Ch_{1,x} &=& -h_{0,T}  \,, \\
    -C\Psi_{1,xY}^-|_0+Gh_{1,x}  &=& -\Psi^-_{0,TY}|_0+Gh_{0,x}-
C\Psi_{0,xy}^+|_0-U_{0,y}\Psi_{0,x}|_0 \,,
\end{eqnarray*}
or
\begin{eqnarray*}
\tilde{\phi}_1-C\tilde{h}_1 &=& -\frac{1}{iK}A_{2,T}\tilde{h}_0 \,, \\
-C\tilde{\phi}_1+(g/k)\tilde{h}_1 &=&
-\frac{1}{iK}A_{2,T}\tilde{\phi}_0^-
+A_2\big{(}(G/K)-U_{,y}C(1/k)\big{)}\tilde{h}_o-
\frac{1}{k}C\big{(}A_1\phi_{a,y}+B_{1-}\phi_{b,y}\big{)} \,.
\end{eqnarray*}

\noindent
Set $V_1=[\tilde{h}_1,\tilde{\phi}_1 ]^T$ and
%\[V_1=
%\left[
%\begin{array}{c} \tilde{h}_1 \\
%\tilde{\phi}_1
%\end{array}
%\right]
%\mbox{ and }
\[
W_1=
\left[ \begin{array}{c} 0 \\
(G/K)-U_{,y}C/K
\end{array}\right], \quad
W_2=
\left[ \begin{array}{c} 0 \\
-C/K \phi_{a,y} \end{array}\right], \quad
W_3=
\left[ \begin{array}{c} 0 \\
-C/K\phi_{b,y}
\end{array}\right] \,;
\]
then the previous set of equations can be written as
\[
MV_1=-\frac{1}{iK}A_{2,T}V_0+A_2W_1+A_1W_2+B_{1-}W_3 \,.
\]
Let $V^T=[C,1]$. $V^T$ is chosen such that $V^TM=0$. Then multiplying the
previous equation with $V^T$ yields
\begin{equation}
\frac{1}{iK}A_{2T} \langle V^TV_o \rangle=
A_2\langle V^TW_1 \rangle+
A_1\langle V^TW_2 \rangle+
B_{1-}\langle V^TW_3 \rangle \,,
\label{Ampr}
\end{equation}
where

\[
\langle V^TV_o \rangle=C\tilde{h_o}+\tilde{\phi_o}=2C\tilde{h_o}=2C \,,
\]
\[
\langle V^TW_1 \rangle=(G/K)-U_{,y}C/K \,,
\]
\[
\langle V^TW_2 \rangle=-C/K\phi_{a,y} \,,
\]
\[
\langle V^TW_3 \rangle=-C/K\phi_{b,y} \,,
\]
where with no loss of generality we have set $\tilde{h}_o=1$;
here $\phi_a,\phi_{a,y}, \dots$ refers to the
values of these quantities at the interface.

Equation (\ref{Ampr}) is the amplitude equation. What is left to do is to
recall the properties of $A_2,A_1^{\pm},B_1^{\pm}$ in order to cast the
amplitude equation into a more convenient form. From the boundary
condition at infinity, we will obtain a condition of the form $B_{1+}=\mu
A_{1+}$, e.g., only one linear combination of the two Frobenius solutions
will give a solution that dies at infinity.
  From equations (\ref{A1}) and (\ref{phaseB}) we have
$A_{1+}=A_{1-}=A_1$,
$B_{1-}=B_{1+}-J$,
and from equation (\ref{b13}) we obtain
$-CA_2+(A_1\phi_a+B_{1-}\phi_b)=0$. Using these results, we can write

\begin{equation}
A_1=\frac{C}{\phi_a+\mu \phi_b}A_2+
\frac{\phi_b}{\phi_a+\mu \phi_b}J \,,
\label{AmpR1}
\end{equation}

\begin{equation}
B_{1-}=\frac{\mu C}{\phi_a+\mu\phi_b} A_2-
\frac{\phi_a}{\phi_a+\mu \phi_b} J .
\end{equation}
We can therefore write the amplitude equation as

\begin{equation}
A_{2T}+\frac{i}{2}\left[ (U_{,y}-G/C)+C
\frac{\phi_{a,y}+\mu \phi_{b,y}}{\phi_a+\mu{\phi_b}}
\right]A_2=
i \left[\frac{\phi_b\phi_{a,y}-\phi_a\phi_{b,y}}
{\phi_a+\mu{\phi_b}} \right] J \quad .
\label{AmpR2}
\end{equation}
%or in a simpler notation
%\begin{equation}
%A_{2,T}+i{\mathcal{C}}_1A_2=i{\mathcal{C}}_2J \qquad
%{\mathcal{C}}_1,{\mathcal{C}}_2 \in \Re
%\end{equation}

%%%%%%%%%%%%%%%%%%%%%%%%%%%%%%%%%%%%%%%%%%%%%%%%%%%%%%%%%%%%%%%%%%%%

\subsection{Strong gravitation and arbitrary density ratio:
$G\gg 1$ and $0<r<1$}

We are now going to focus on cases for which $G \gg 1$, which are
associated with small growth rate. As before, we use the time
scaling $\partial_t=\epsilon\partial_T-C\partial_x$, where
$\epsilon$ is of the order of the linear growth rate (to be defined
below). The governing equation is then equation (\ref{Navier}), with
the boundary condition now given by

\begin{equation}
\epsilon h_{,T}-Ch_{,x}+\Psi^+_{,x} =0 \,,
\label{b21}
\end{equation}
\begin{equation}
\epsilon h_{,T}-Ch_{,x}+\Psi^-_{,x} =0 \,,
\label{b22}
\end{equation}

\begin{equation}
r\Big{[}  \epsilon \Psi_{,Ty}^+
-C\Psi^+_{,xy}-U^+_{,y}\Psi^+_{,x} \Big{]}
-\Big{[}\epsilon \Psi_{,Ty}^--C\Psi^-_{,xy}\Big{]}
-Gh_{,x}(1-r) = 0 \,.
\label{b23}
\end{equation}
We expand the surface elevation and stream function as:
\begin{equation}
\Psi=\epsilon^2\Psi_0+\epsilon^3\Psi_1 + \dots
\end{equation}
\begin{equation}
h=\epsilon^2h_0+\epsilon^3h_1 + \dots
\end{equation}
To zeroth order we then have again

\[
(U-C) \nabla^2\Psi_{0,x}-U_{yy}\Psi_{0,x}=0 \,;
\]
expanding in normal modes, $\Psi_0=A_2\phi(y)e^{iKx}+ \mbox{c.c.}$, we
have

\[
\phi_{,yy}-\left[K^2-\frac{U_{,yy}}{U-C}\right]\phi=0 \,,
\]
where we have again focused on the most unstable mode. Since we are dealing
with the large $G$ case, we know from equation (\ref{BOUND})
that the wave-numbers of the unstable modes must satisfy (in the limit of
large $G$)

\begin{equation}
K \ge \frac{1-r}{1+r}G={\mathcal{A}}_tG \gg 1
\label{boundG}
\end{equation}
(where ${\mathcal{A}}_t$ is the Atwood number); we can then use the
results from the WKBJ expansion discussed in Appendix A. These results
show that the value of $\Psi_0$ and $\Psi_{0,y}$ at the interface is

\begin{equation}
\Psi_0|_0=\left[A_1(\sin(I_1)e^{-I_2}+\cos(I_1)e^{+I_2})-
\frac{1}{\pi}J\sin(I_1)e^{+I_2} \right]e^{iK(x-Ct)}+ \mbox{c.c.}
\label{psi0W}
\end{equation}

\begin{equation}
\Psi_{0,y}|_0=\left[A_1(\sin(I_1)e^{-I_2}-\cos(I_1)e^{+I_2})+
\frac{1}{\pi}J\sin(I_1)e^{+I_2} \right]e^{iK(x-Ct)}+ \mbox{c.c.} \,,
\label{phiy0W}
\end{equation}
where $I_1 \sim 1/K$ and $I_2 \sim K\ge {\mathcal{A}}_tG \gg 1$
  are defined in the Appendix.
Unlike the linear case, we assumed that the phase change is $J$, defined
as in equation (\ref{J}) in the previous section (\S4.1.2). We avoid
repeating the inner scale calculations since they are identical to
that for the small density ratio case. The slow time scale is now defined
by the value of $C_i$, which is exponentially small and is given by
expression (\ref{exp}). We will therefore define $\epsilon \equiv
e^{-2I_2}$. Equations (\ref{psi0W}) and (\ref{phiy0W}) can then be
written as

\begin{equation}
\Psi_0|_0=\left[A_1\cos(I_1)-\frac{1}{\pi}J\sin(I_1)+\epsilon
A_1\sin(I_1)
\right]e^{iKx}+ \mbox{c.c.} \,,
\label{psiW}
\end{equation}
and
\begin{equation}
\Psi_{0,y}|_0=-K\left[A_1\cos(I_1)-\frac{1}{\pi}J\sin(I_1)-\epsilon
A_1 \sin(G_1)\right]e^{iKx}+ \mbox{c.c.} \,,
\label{psiyW}
\end{equation}
where we have rescaled $\Psi_0$ so that it is of order unity at the
interface.

Writing the zeroth order stream function $\Psi_0^-$ below the
interface, and the surface elevation $h_0$ as

\begin{equation}
\Psi_0^-=A_3e^{iKx+Ky}+ \mbox{c.c.} \,,
\end{equation}
\begin{equation}
h_0=A_2e^{iKx}+ \mbox{c.c.}\,,
\end{equation}
the first order boundary conditions give us
\begin{equation}
M\cdot V_0 =0 \,,
\end{equation}
where

\begin{equation}
M= \left[
\begin{array}{ccc}
-C & 1 & 0 \\
-C & 0                          & 1\\
-G(1-r) &rCK&CK \\
\end{array}
\right]
\,,~~
V_0=\left[
\begin{array}{c}
A_2 \\
\Phi  \\
A_3 \\
\end{array}
\right] \,,
\label{M}
\end{equation}
and $\Phi=(2A_2\cos(I_1)-\frac{1}{\pi}J\sin(I_1))$ is the amplitude of
the stream function at the interface. Here we have discarded the $U_y$
term since it is of order $1/K$.
For a non-trivial solution, we must have det$(M)=0$.
This leads us to the relations
\begin{equation}
C=\sqrt{G{\mathcal A}_t/K}
\end{equation}
and
\begin{eqnarray}
A_2C &=& \Phi \,, \\
A_2C &=& A_3 \,.
\end{eqnarray}
At the next order, we have
\[
(U-C) \nabla^2\Psi_{1,x}-U_{,yy}\Psi_{1,x}=\nabla^2\Psi_{0,T} \,,
\]
and by letting $\Psi_1=\hat{A}_1e^{iKx}\phi_1(y)$, we have
\begin{equation}
\hat{A}_1\left[\phi_{1,yy}-\left(K^2+\frac{U_{,yy}}{U-C}\right)\phi_1\right]
=\hat{A}_{1,T}\frac{U_{,yy}}{U-C}\phi_0 \,.
\label{secord}
\end{equation}
Since we are still in the large $K$ limit, we can still use the WKBJ
approximation for the above non-homogeneous equation. One needs to notice
that the non-homogeneous term is going to be of order $1/K^2$ everywhere
except close to the critical layer, and that special care is needed to be
taken there. Fortunately, the exact value of $\phi_1$ at the interface is
not needed --- it is sufficient to know that the solution of
equation (\ref{secord}) below and away from the critical layer is

\begin{equation}
\phi_1=
\hat{A}_{1a}\frac{1}{w}e^{+K\int_0^{y_c}wdy'}e^{-Ky}+
\hat{A}_{1b}\frac{1}{w}e^{-K\int_0^{y_c}wdy'}e^{+Ky} \,,
\label{tricky}
\end{equation}
where $w=(K^2-U_{,yy}/(U-C))^{1/2}$, and where the values of
$\hat{A}_{1a},\hat{A}_{1b}$ depend on $A_{1T}$, and can be obtained by
matching with the inner solution at the critical layer. The second term
in equation (\ref{tricky}) is exponentially small (order
$\epsilon$), so that at the interface we have $\phi_{1,y}=-K\phi_1$; this
suffices for the remainder of our analysis. The stream function of
the lower fluid $\Psi_1^-$ and the elevation $h_1$ at this order are
\[
\Psi_1^-=\hat{A}_3e^{iKx+Ky}+ \mbox{c.c.} \,,
\]
\[
h_1=\hat{A}_2e^{iKx}+ \mbox{c.c.}
\]
The boundary conditions at the interface are

\begin{equation}
M \cdot V_1=-\frac{1}{iK}\partial_T W_1+W_2 \,,
\label{VEC2}
\end{equation}
(where $M$ was defined in equation \ref{M})

\begin{equation}
V_1= \left[
\begin{array}{c}
\hat{A}_2 \\
\phi_1 \hat{A}_1
\\ \hat{A}_3 \\
\end{array}
\right],
W_1=\left[
\begin{array}{c}
A_2 \\
A_2 \\
-K(r\Phi+A_3)
\end{array}
\right] =A_2 \left[
\begin{array}{c}
1 \\
1 \\
-(r+1)KC
\end{array}
\right] \,,
\end{equation}
and
\begin{equation}
W_2=\left[
\begin{array}{c}
-\sin(I_1)A_1 \\
    0\\
rKC\sin (I_1)A_1
\end{array}
\right] \,.
\end{equation}
As before, let $V^T=[-rCK,-CK,1]$ where $V^T$ is such
that $V^T M=0$. Multiplying equation (\ref{VEC2}) with $V^T$, we obtain

\[
V^TW_1=-2(1+r)CKA_2, \quad \quad V^TW_2=2rCKA_2\sin(I_1) \,.
\]
Hence, we arrive at the amplitude equation for large $G$

\begin{equation}
\frac{1}{iK}V^TW_{1,T}=V^TW_2 \,,
\end{equation}
or
\begin{equation}
A_{2,T}=-iK\frac{r}{1+r}\sin(I_1)A_1 \,,
\label{AmpG1}
\end{equation}
with
\begin{equation}
CA_2=2A_1\cos(I_1)-J\frac{1}{\pi}\sin(I_1) \,.
\label{AmpG2}
\end{equation}
The amplitude evolution can now be determined from equations
(\ref{AmpG1})-(\ref{AmpG2}).

%%%%%%%%%%%%%%%%%%%%%%%%%%%%%%%%%%%%%%%%%%%%%%%%%%%%%%%%%%%%%%%%%%%%%%%%%
%%%%%%%%%%%%%%%%%%%%%%%%%%%%%%%%%%%%%%%%%%%%%%%%%%%%%%%%%%%%%%%%%%%%%%%%%

\section{Results}

\subsection{Preliminaries}
\label{prelim}

For both cases we have considered, the amplitude equations to be
solved in order to determine the nonlinear evolution of the unstable
mode  can be written in to the following more general form:

\begin{equation}
    A_{2,T}+i{\mathcal{C}}_1A_2=i{\mathcal{C}}_2J
\label{A2T}
\end{equation}

\begin{equation}
A_2= {\mathcal D}_1 A_1 +{\mathcal D}_2 J
\label{A21}
\end{equation}

\begin{equation}
J=\frac{K}{2\pi}\frac{U'_c}{U''_c}\int^{+\pi/K}_{-\pi/K}e^{-iKx}
\int^{+\infty}_{-\infty}Z \quad dYdx
\equiv \langle e^{-iKx}Z \rangle
\label{J1}
\end{equation}

\begin{equation}
Z_{,T}+U_c'YZ_{,x}-\tilde{\Psi}_{0,x}Z_{,Y}-\nu
Z_{,YY}=U_c''\tilde{\Psi}_{0,x}
\label{ZT1}
\end{equation}
with

\[
\lim_{Y\to\infty}Z=\left[ A_1\frac{U_c''}{2U_c'}\frac{1}{Y}+
\frac{U_c''}{2U_c'^2}\frac{1}{Y^2}A_{1,T}\right] e^{iKx}+c.c. \]
and

\[ \tilde{\Psi}_0=A_{1}e^{iKx}+A_{1}^* e^{-iKx} \]

The equations above should be interpreted the following way. Equation
(\ref{A21}) expresses the continuity of the normal velocity at the
interface: It imposes the constraint that the phase and amplitude of the
perturbation of the surface wave (given by $A_2$) is the same as the
perturbation of the wind (given by ${\mathcal D}_1 A_1 +{\mathcal D}_2 J$)
including the component that comes from the phase change at the interface.
Equation (\ref{A2T}) is Newton's law, or alternatively can be viewed as a
statement about the continuity of pressure at the interface; it gives
the growth of the amplitude $A_2$ due to the out of phase component of the
pressure.  Equation (\ref{ZT1}) gives the evolution of the vorticity inside
the critical layer that determines the phase change, and involves the
non-linear term $\tilde{\Psi}_{0,x}Z_{,Y}$. The coefficients ${\mathcal
C}_1,{\mathcal{C}}_2,{\mathcal D}_1, {\mathcal D}_2, \in \Re$ are obtained
from the linear theory, and are given in equations (\ref{AmpR1}),
(\ref{AmpR2}), (\ref{AmpG1}) and (\ref{AmpG2}). ${\mathcal
C}_1$ and ${\mathcal{D}}_2$ are terms that involve small correction to the
real part of $C$ due to the gradient of the velocity at the interface
$U_{,y}$ and due to the part of the pressure perturbation that is in phase
with the traveling wave. ${\mathcal C}_2$ and ${\mathcal{D}}_1$ involve
the part of the pressure perturbation that is out of phase with the
traveling wave; their sign determines the nature of the instability.

Dropping the non-linear term in
eq. (\ref{ZT1}) we obtain $J=i\pi A_1$ in the linear case\citep{Drazin81}. 
Assuming an exponential growth rate $A_i=e^{\gamma T} a_i$, 
the above equation can be written as:

\begin{eqnarray}
\pi {\mathcal{C}}_2a_1 & +  (\gamma + i{\mathcal{C}}_1)a_2 &=  0 \,, \\
({\mathcal{D}}_1- i\pi{\mathcal{D}}_2)a_1 & - a_2 &=  0 \,;
\end{eqnarray}
the growth rate then is given by

\begin{equation}
\gamma=-\frac{\pi {\mathcal{C}}_2 {\mathcal{D}}_1}
{{\mathcal{D}}_1^2+\pi^2{\mathcal{D}}_2^2}
-i\left(\frac{\pi^2{\mathcal{D}}_2{\mathcal{C}}_2}
{{\mathcal{D}}_1^2+\pi^2{\mathcal{D}}_2^2}+{\mathcal{C}}_1\right) \,.
\end{equation}
For a positive growth rate we therefore need to have
${\mathcal{C}}_2 {\mathcal{D}}_1 < 0$.

We note as an aside that there are several conservation laws at work
here.\footnote{In the equations (\ref{momentum}), (\ref{enstrophy1}), and
(\ref{energy}), the integration over $x$ was assumed to be taken first.}
First, the vorticity is conserved inside the critical layer
$\langle Z \rangle_{,T}=0$, which implies $\langle Z \rangle=0$ since the
initial $Z$ had infinitesimal amplitude.
Second, by noting that $\langle \Psi_{0,x}Z \rangle= iK(J^*A_1-JA_1^*)$, one
can show that the following laws hold:

\begin{equation}
\frac{d}{dT}\left\{
K|A_2|^2-{\mathcal{C}}_2{\mathcal{D}}_1\langle YZ \rangle
\right\}=0 \,;
\label{momentum}
\end{equation}
\begin{equation}
\frac{d}{dT}\left\{
KU_c''|A_2|^2+\frac{1}{2}{\mathcal{C}}_2{\mathcal{D}}_1
\langle Z^2 \rangle \right\}=
-{\mathcal{C}}_2{\mathcal{D}}_1\nu \langle Z_{,Y}^2 \rangle \,;
\label{enstrophy1}
\end{equation}
\begin{equation}
\frac{d}{dT}\left\{
{\mathcal D}_1 {\mathcal C}_2
\langle (\tilde{\Psi}_0+\frac{1}{2}U'_cY^2)Z\rangle +
{\mathcal D}_2 {\mathcal C}_2 |J|^2-{\mathcal C}_1|A_2|^2
\right\}=\nu U'{\mathcal C}_2\langle Z\rangle \,.
\label{energy}
\end{equation}
Recalling that the velocity inside the critical layer is given by
$[u-C,w]=[\epsilon U'_cY+ \epsilon^2(1/2U_c''Y^2+\Psi_{2,Y})+\dots ,
\epsilon^2 \Psi_{0,x}+\dots ]$
we can identify the first relation (eq.\ \ref{momentum}) to correspond to
the conservation of momentum. Combining equations (\ref{momentum}) and
(\ref{enstrophy1}), we obtain the conservation of enstrophy inside the
critical layer,

\begin{equation}
\frac{d}{dT}\langle (U''_cY+Z)^2 \rangle = -\nu \langle Z_{,Y}^2 \rangle
\,.
\label{enstrophy}
\end{equation}
The third equation (\ref{energy}) can be regarded as a statement of the
conservation of energy.

%%%%%%%%%%%%%%%%%%%%%%%%%%%%%%%%%%%%%%%%%%%%%%%%%%%%%%%%%%%%%%%%%%%%%%%%%%
%%%%%%%%%%%%%%%%%%%%%%%%%%%%%%%%%%%%%%%%%%%%%%%%%%%%%%%%%%%%%%%%%%%%%%%%%%

\subsection{Quasi-steady state}
\label{quasi}

%%%%%%%%%%%%%%%%%%%%%%%%%%%%%%%%%%%%%%%%%%%%%%%%%%%%%%%%%%%%%%%%%%%%%%%%%

\begin{figure}[!htb]
\centerline{\epsfysize = 3.in \epsffile{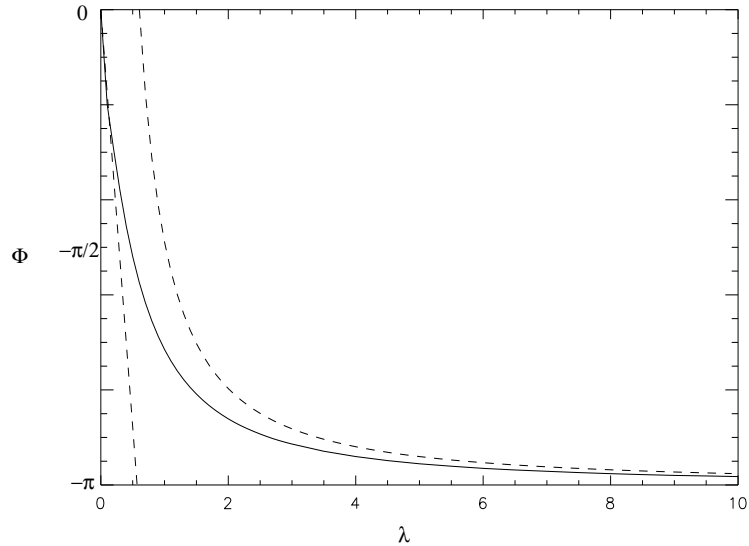}}
\caption{The Haberman function}
\label{Habberf}
\end{figure}

An interesting limit in the set of our equations is when the rescaled
viscosity is
large enough to play a dominant role inside the critical layer. With this
assumption we can drop the time derivative term in equation (\ref{Z});
the functional form of $Z$ then just depends on the value of $A_1$, which
then makes $J$ just a function of $A_1$. More specifically we have

\begin{equation}
\nu Z_{,YY}+\tilde{\Psi}_{0,x}Z_{,Y}-U_c'YZ_{,x}=
-U_c''\tilde{\Psi}_{0,x} \,;
\end{equation}
by letting  $A_1=R(T)e^{i\Theta(T)}$ and
$\xi=Kx+\Theta$, we have

\begin{equation}
\nu Z_{,YY}-2RK\sin(\xi)Z_{,Y}-U_c'KYZ_{,\xi}=2U_c''RK\sin(\xi) \,.
\end{equation}
Let $Y=\sqrt{2}\eta(R/U_c')^{1/2}$,
$Z=\sqrt{R/2U_c'}U_c''\hat{Z}(\xi,\eta)$ and
$\lambda=\nu\sqrt{U_c'}/(K(2R)^{3/2})$; we then obtain

\begin{equation}
\lambda\hat{Z}_{,\eta\eta}-\sin(\xi)\hat{Z}_{,\eta}-
\eta \hat{Z}_{,\xi}=2\sin(\xi) \,;
\end{equation}
Equation (\ref{J}) then becomes
\[
J=\frac{K}{2\pi}\frac{U'_c}{U''_c}\int^{+\infty}_{-\infty}
\int^{+\pi/K}_{-\pi/K}
e^{-iKx}ZdxdY=
\frac{1}{2\pi }R \cdot e^{i\Theta}
\int^{+\infty}_{-\infty}
\int^{+\pi+\Theta}_{-\pi+\Theta}
e^{-i\xi}\hat{Z}d\xi d\eta
\]

\[
=\frac{1}{2\pi}R \cdot e^{i\Theta}
\left[ \int^{+\infty}_{-\infty}
\int^{+\pi}_{-\pi}
\cos\xi\hat{Z}d\xi d\eta
-i \int^{+\infty}_{-\infty}
\int^{+\pi}_{-\pi}
\sin\xi\hat{Z}d\xi d\eta \right]
\]

\[
=-i \frac{1}{2\pi }A_1
\int^{+\infty}_{-\infty}
\int^{+\pi}_{-\pi}
\sin\xi\hat{Z}d\xi d\eta
\]
or
\[J=-i A_1 \Phi_1(\lambda) \]
with
\[
\Phi_1(\lambda)=\frac{1}{2\pi}\int^{+\pi}_{-\pi}\int^{+\infty}_{-\infty}
\sin\xi\hat{Z}d\eta d\xi.
\]
$\Phi_1(\lambda)$ was first studied numerically by 
\cite{Haberman72}. Its values range from $-\pi$ for $\lambda \to \infty$
to 0 for $\lambda \to 0$. Its asymptotics for $\lambda \to 0$ and $\lambda
\to \infty$ are given by
\begin{equation}
\Phi(\lambda)= \left\{
\begin{array}{cr}
-a_1\lambda+a_2\lambda^{3/2}+O(\lambda^2)
\quad & \lambda \ll 1 \\
-\pi+ b_1 \lambda^{-4/3} +O(\lambda^{-8/3})
\quad & \lambda \gg 1
\end{array}
\right.
\end{equation}
where $a_1=5.5151 \dots$, $a_2=4.2876\dots$ and $b_1=1.6057\dots$ . 
The derivation for the above asymptotics can be found 
in \citep{Churilov96}. We illustrate the behavior of $\Phi(\lambda)$ 
in figure.~(\ref{Habberf}),
constructed by combining numerical evaluation with the aforementioned
asymptotics.

The amplitude equation then can be written as
\[ A_{2,T}+i{\mathcal{C}}_1A_2={\mathcal{C}}_2
A_1\Phi\left(\frac{\nu\sqrt{U_c'}}{K(2|A_1|)^{3/2}}\right)\] \,,
with
\[A_2={\mathcal{D}}_1A_1+i{\mathcal{D}}_2
A_1\Phi\left(\frac{\nu\sqrt{U_c'}}{K(2|A_1|)^{3/2}}\right) \,.
\]

We have thus ended up with an ordinary differential equation for the
amplitude of the wave. By setting $A_1=R(T)exp[i\Theta(T)-i{\mathcal
C}_1T]$, and after some algebra, one can show that

\begin{equation}
\left(
{\mathcal D}_1^2+{\mathcal D}_2^2\Phi\tilde{\Phi}\right) R_{,T}=
{\mathcal D}_1{\mathcal C}_2R^2\Phi \,,
\end{equation}

\begin{equation}
\left(
{\mathcal D}_1^2+{\mathcal D}_2^2\Phi\tilde{\Phi}\right) \Theta_{,T}=
{\mathcal C}_2{\mathcal D}_2 \Phi\tilde{\Phi} \,,
\end{equation}
where $\tilde{\Phi}=(R\Phi)_{,R}$. For large $R$ at late times we can
conclude that

\begin{equation}
|A_2|\sim |A_1| \equiv R \sim \nu T^{2/3}
\label{AmQT}
\end{equation}
and

\begin{equation}
\Theta\sim \nu^2 T^{-1} \,.
\label{ThetaT}
\end{equation}

This is one of the basic results of our calculations. As we are going to
show later, this result becomes valid for later times even for cases for
which the rescaled viscosity is smaller than one. An important implication
of this result is that the amplitude grows with an algebraic power instead
of the initial exponential variation. Another important feature is that
the growth rate linearly depends on the viscosity, unlike the case in linear
theory. (In linear theory, a weak viscosity was giving the same phase change
``$-i\pi$", but 
the resulting growth rate was independent of $\nu$.)
Finally we note that the phase of the waves $\Theta$,
goes assymptoticaly to zero at late times according to equation (\ref{ThetaT}).

%%%%%%%%%%%%%%%%%%%%%%%%%%%%%%%%%%%%%%%%%%%%%%%%%%%%%%%%%%%%%%%%%%%%%%%%%%%
%%%%%%%%%%%%%%%%%%%%%%%%%%%%%%%%%%%%%%%%%%%%%%%%%%%%%%%%%%%%%%%%%%%%%%%%%%%

\subsection{Numerical Results}
\label{nume}

Next we investigate the weakly non-linear evolution of the wave
by solving the set of equations (\ref{A2T}-\ref{ZT1}) numerically. To
solve the advection equation (\ref{ZT1}), we used a code which is spectral
in $x$ and finite difference in $Y$. The domain we used was $(-50,50)$ in
$Y$ and $(0,2\pi)$ in $x$. Up to 1024 grid points were used in the $Y$
direction and 63 modes were kept in $x$. The boundary conditions were
satisfied to order $1/Y$, although the asymptotic behavior of $Z$ was
taken into account when we evaluated the integral in equation (\ref{J1}).
The code was tested by comparing with a fully pseudo-spectral code as
well as with already published results.

Before we begin solving the amplitude equations, however, we re-scale our
system so that we are left with a minimum number of free parameters. As
in the previous section, by letting $Z=U''/\sqrt{U'}\tilde{Z}$,
$Y=\eta/\sqrt{U'}$, $T=\tau/(\sqrt{U'}k)$, $\xi=Kx$ and
$\nu=\nu'/\sqrt{U'}$ we obtain the following equation to be solved:
\[
\tilde{Z}_{,\tau}+ \eta \tilde{Z}_{,\xi}-
\tilde{\Psi}_{0,\xi}\tilde{Z}_{,\eta}
-\nu'\tilde{Z}_{,\eta \eta}=
\tilde{\Psi}_{0,\xi}  \,,
\]
with
\[
J=\frac{1}{2\pi}\int^{+\pi}_{-\pi}\int^{+\infty}_{-\infty}
e^{-i\xi}\tilde{Z}d\eta d\xi \,.
\]
Furthermore, by rescaling $A_1$ to $|{\mathcal D}_1|^2/|{\mathcal C}_2|^2
A_1$, and $A_2$ to $|{\mathcal D}_1|/|{\mathcal C}_2|^2 A_2$, we can
always scale our system so that ${\mathcal D}_1=1$ and ${\mathcal
C}_2=-1$. Finally, the coefficient ${\mathcal C}_1$ can always be set to
zero by performing a Galilean transformation $(\tau \to \tau + {\mathcal
C}_1\xi)$ and shifting the critical layer by $Y \to Y-{\mathcal C}_1$.
The last transformation corrects the position of the critical layer to
order $\epsilon$, and justifies the neglect of the corresponding terms in
\S4.2. We are left therefore with one independent parameter,
${\mathcal D}_2$, to investigate; this parameter is a measure of the 
feedback of the gravity wave to the critical layer (${\mathcal D}_2=0$
gives the ``free'' evolution of a critical layer uncoupled from gravity
waves). In the following, we examine five cases, for five values of
${\mathcal D}_2$.

\begin{figure}[!htb]
\centerline{\epsfysize = 1.4in \epsffile{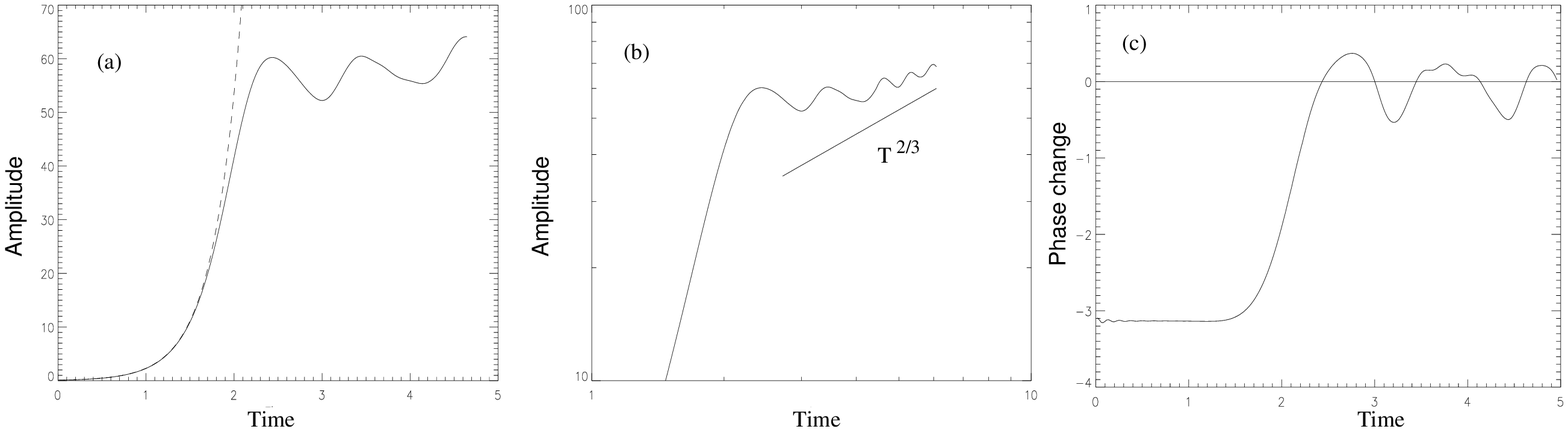}}
\caption{The evolution of the wave amplitude $|A_1|$ (panels [a] and [b]),
and of the phase change across the critical layer (panel [c]), as a function
of time. The dashed line panel [a] gives the linear prediction; the solid
line in panel [b] is intended to illustrate $T^{2/3}$ scaling.}
\label{A4_1}
\end{figure}

\begin{figure}[!htb]
\centerline{\epsfysize = 1.2in \epsffile{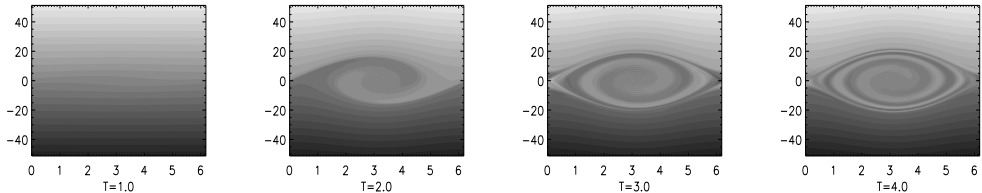}}
\caption{Time sequence of the evolution of the vorticity inside
the critical layer for ${\mathcal D}_2=0$, using a grey-scale
representation for the vorticity.}
\label{V4_1}
\end{figure}

We first examine ${\mathcal D}_2=0$ case, e.g.
the evolution of a `free critical layer'.  In figure~(\ref{A4_1}a) we
plot the amplitude as a function of time; the dashed line gives the linear
prediction. The amplitude grows exponentially for early times, but as
non-linearities become important the growth rate decreases, and an
oscillatory behavior begins. figure~(\ref{A4_1}b) shows the same thing,
but plotted on a log-log scale, and compared with the $T^{2/3}$ scaling; we
note that the agreement is good for later times, although the persistence
of the oscillations indicate that viscosity is not yet fully dominant.
Figure~(\ref{A4_1}c) gives us more information about the system; here we
plot the phase change $\Im \{-J/A_1\}$ as a function of time. The phase
change (which is responsible for the instability) remains constant, and
equal to $-\pi$, in the linear regime; in the non-linear regime it drops to
zero and then oscillates around this value. In figure~(\ref{V4_1}) we display
the evolution of the total vorticity $\tilde{Z}+\eta$ inside the critical
layer, using a grey-scale representation; time frame were taken every $0.5$
time units (the first panel is at $T=1$). The vorticity of the mode
rises,
as predicted by the linear theory, by extracting vorticity from the mean
flow. (From eqs.~(\ref{enstrophy1}) and (\ref{enstrophy}) we know that the
total vorticity and enstrophy are conserved; therefore the increase of
$Z^2$ must lead to a decrease of $ZY$.)\ \ As the non-linear term becomes
important, the mean flow advects the vorticity in the vertical direction,
leading to the `winding up' motion seen in the panels with $T \ge 2$,
forming the cat's eyes pattern. The imaginary part of the integral $J$
(eq.~(\ref{J1})) expresses the excess of vorticity of the right side of
each panel, from the left. After the first turnover, the vorticity is
redistributed inside the critical layer more evenly, leading to smaller
values of $J$. Comparing figure~(\ref{A4_1}) with figure~(\ref{V4_1}), it
can be seen that deviations from linearity and the $-i\pi$ phase change
start when the vertical motion begins, and the phase change has dropped to
zero when the first full turn over has been completed. At later times and
after a few turnovers, small scale structure has been generated in the form
of a `twisted' vortex sheet with sharp boundaries. As this procedure
carries on, at some point in time, viscosity (now matter how small) will
become important, and the advection terms in eq.~(\ref{ZT1}) will be
balanced by the diffusion of vorticity; at that point, the asymptotic
behavior predicted by the quasi-steady state for large times can been shown
to hold.

\begin{figure}[!htb]
\centerline{\epsfysize = 1.2in \epsffile{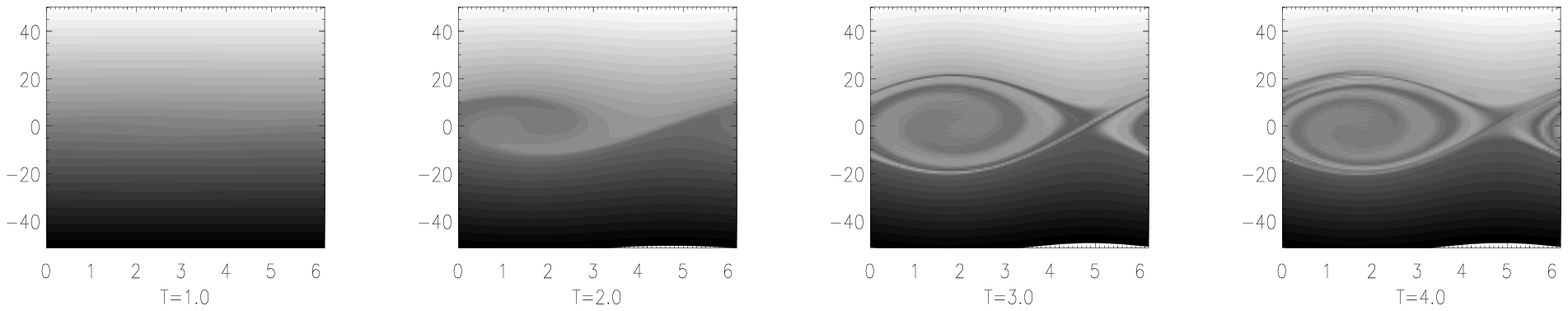}}
\caption{As in figure~(\ref{V4_1}), but for ${\mathcal D}_2=0.1$}.
\label{V4_8}
\end{figure}
\begin{figure}[!htb]
\centerline{\epsfysize = 1.2in \epsffile{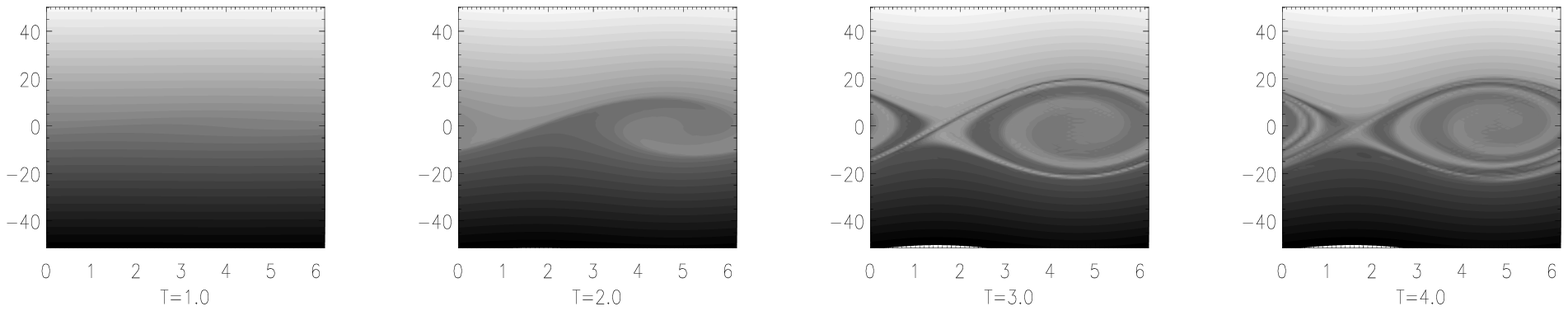}}
\caption{As in figure~(\ref{V4_1}), but for ${\mathcal D}_2=-0.1$}
\label{V4_9}
\end{figure}

Next we investigate the effect of non-zero ${\mathcal D}_2$. The cases we
will examine will be for ${\mathcal D}_2= \pm 0.1$ and $\pm 0.3$. We start
with the two cases ${\mathcal D}_2= \pm 0.1$; here we note that for
${\mathcal D}_2= - 0.1$, linear theory predicts a right traveling wave, but
for ${\mathcal D}_2= 0.1$ predicts a left traveling wave. In
figures~(\ref{V4_8}) and (\ref{V4_9}) we show grey-scale representations
of the evolution of vorticity for these two cases. We see that the result is
two traveling vortices, as predicted  by the linear theory, but the phase
velocity is decreasing with time and the vortices seem to stop at a
distance $\pm \pi/2$ from their initial position. This is in agreement with
the quasi-steady limit, which predicts that the phase $\Theta(T)$ in
equation (\ref{ThetaT}) should decrease as  $T^{-1}$. Small scale
differences appear in the vorticity field when compared to
figure~(\ref{V4_1}), but these seem to be minor.

\begin{figure}[!htb]
\centerline{\epsfysize = 1.2in \epsffile{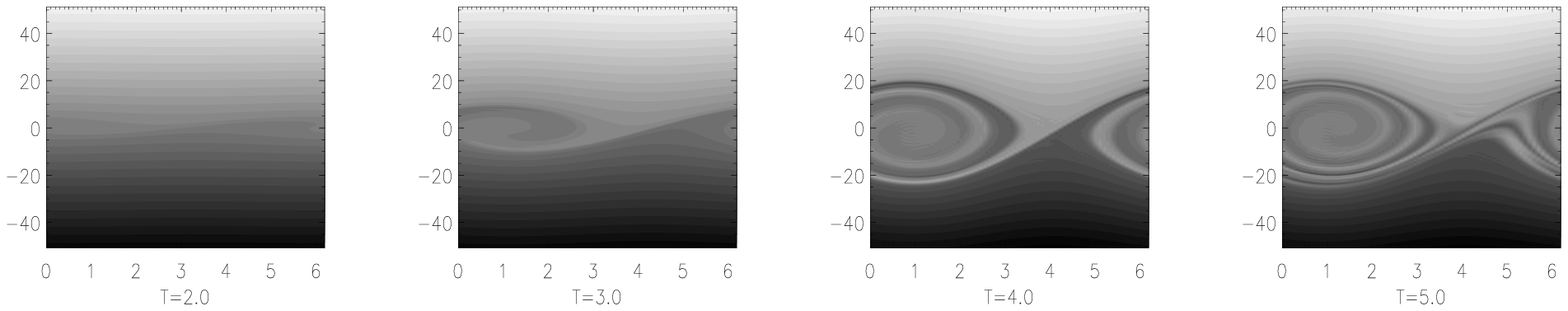}}
\caption{As in figure~(\ref{V4_1}), but for ${\mathcal D}_2=0.3$}
\label{V4_10}
\end{figure}
\begin{figure}[!htb]
\centerline{\epsfysize = 1.2in \epsffile{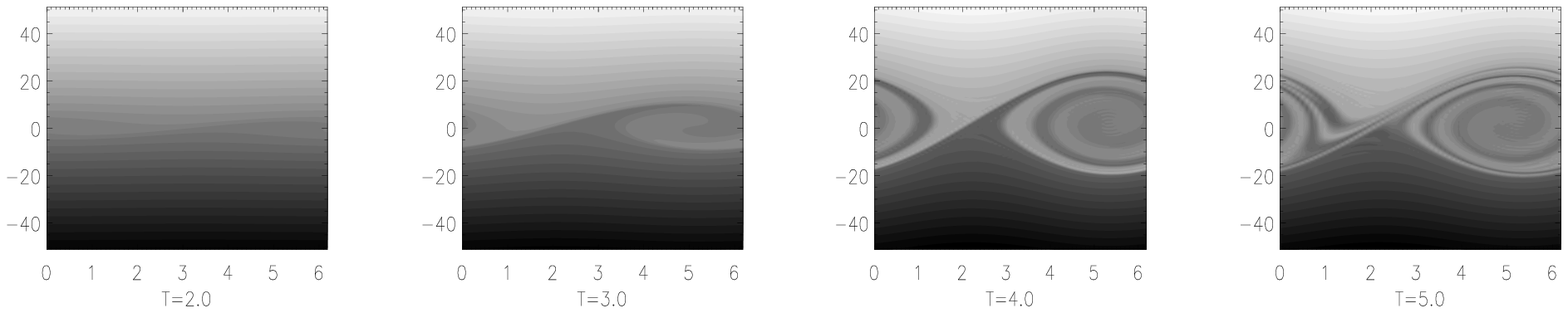}}
\caption{As in figure~(\ref{V4_1}), but for ${\mathcal D}_2=-0.3$}
\label{V4_11}
\end{figure}

By considering lightly larger values of ${\mathcal D}_2$ ($=\pm0.3$), we
can obtain an indication of how this affects the amplitude of the wave.
In figures.~(\ref{V4_10})-(\ref{V4_11}) we plot again grey-scale
representations of the vorticity for four different times for the ${\mathcal
D}_2=0.3$ and ${\mathcal D}_2=-0.3$ cases, respectively. The linear growth
rate is 1.8 times smaller, so the plots are at later times than the
previous cases shown. Differences in the evolution of the vorticity can be
seen especially at the `saddle point' (the point where the two vortices
meet).

\begin{figure}[!htb]
\centerline{\epsfysize = 1.4in \epsffile{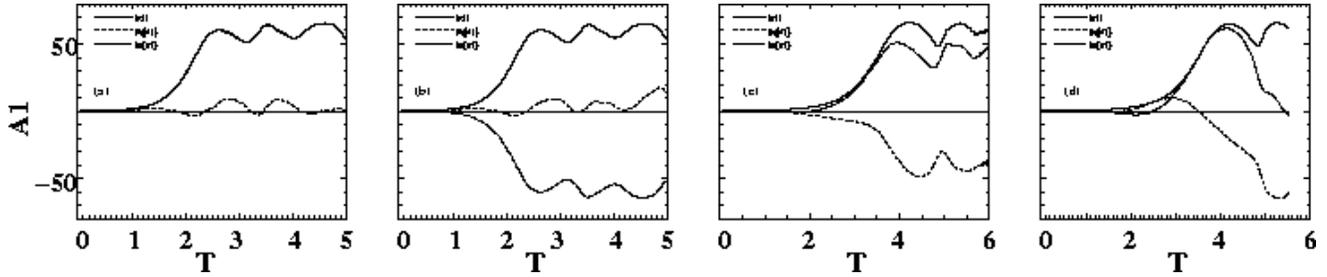}}
\caption{Evolution of the wave amplitude $|A_1|$ for the four cases,
${\mathcal D}_2 = 0.1$ (panel [a]), $-0.1$ (panel [b]),  $0.3$ (panel
[c]), and $-0.3$ (panel [d]). In each panel, the solid lines show the
magnitude of the amplitude, while the two sets of dashed lines show the
evolution of the real and imaginary parts of the amplitude.}
\label{A4_2}
\end{figure}

\begin{figure}[!htb]
\centerline{\epsfysize = 1.4in \epsffile{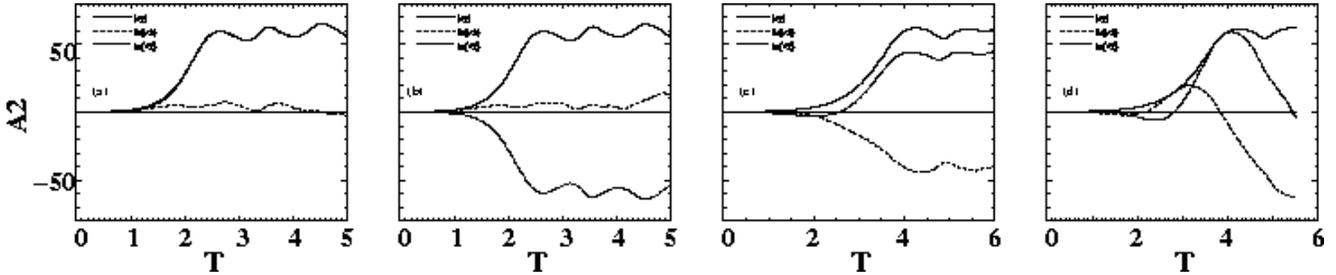}}
\caption{As in figure~(\ref{A4_2}), but for wave amplitude $|A_2|$.}
\label{A3_1}
\end{figure}

A more quantitative appreciation of the time evolution of the wave
amplitude can be obtained by directly plotting the evolution of the
amplitudes $A_1$ and $A_2$ for the four cases of non-zero ${\mathcal D}_2$
we have examined, as shown in figures~(\ref{A4_2}) and (\ref{A3_1}).
In all panels, we show both the (unsigned) wave amplitude magnitude, as well
as the real and imaginary parts of the amplitude; each panel of each figure
corresponds to a different value of ${\mathcal D}_2$, as indicated.

Consider first the cases ${\mathcal D}_2 = \pm 0.1$. The fact that a
positive imaginary part is dominant in  panels [a] of
figures~(\ref{A4_2})-(\ref{A3_1}), and a negative imaginary part in panels
[b] of figures~(\ref{A4_1})-(\ref{A4_2}), just expresses the fact that in this
case, the two vortices moved by a quarter of the wave length and then
stopped. We note that the magnitude of the amplitude at the point where
the oscillations start (e.g., the peak of the `first bump') does not seem
to have been affected by the value of ${\mathcal D}_2$. Now consider
${\mathcal D}_2 = \pm 0.3$. We mentioned just above the appearance of
small-scale differences in the grey-scale representations of the vorticity
field as the magnitude of ${\mathcal D}_2$ is increased; however, the
quantitative plots of the wave amplitude show that these differences in the
small-scale structures do not seem to affect the evolution of the
amplitudes $A_1$ and $A_2$: for example, in panels [c] of figure~(\ref{A4_2}),
we see that the value of $A_1$ at which the first bounce of the amplitude
occurs does not seem to depend strongly on the value of ${\mathcal D}_2$. We
conclude therefore that the basic features of the weakly non-linear
evolution of the unstable mode are not affected by the exact value of
${\mathcal D}_2$, at least in the examined range $-0.3 \leq {\mathcal D}_2
\leq 0.3$.
 
%Finally we note that due to the fact that large values of
%${\mathcal D}_2$ give a small growth rate (for ${\mathcal D}_2=1$ we
%obtain a 10 times smaller growth rate than for the first case),
%computational time are longer. Since the given results indicate no
%strong
%dependence on ${\mathcal D}_2$, no further cases were examined.

%%%%%%%%%%%%%%%%%%%%%%%%__________________%%%%%%%%%%%%%%%%%%%%%%%

\section{Mixing in the cat's eye}

In the previous sections we have examined the cat's
eye vortices in the nonlinear evolution of
the critical layer in the wind coupled to a surface wave.
We have found that the general flow structure to be similar to
those in cases without coupling to the gravity waves.
Such vortical structures are typical of
the weakly nonlinear evolutions of parallel flows \citep{Balmforth01}.
However, their mixing properties may be very different
(depending on the details of the underlying flow) in spite of great
similarity in the general flow features.  We thus investigate
the mixing due to these cat's eyes from the weakly nonlinear
amplitude equations for a few different cases.

In our amplitude equation \ref{ZT1},
the vorticity $Z$ is advected by velocity
$(u,v) = (Y, -\Psi_{0x})$, with $\Psi_0\equiv A_1 e^{iKx}+ \mbox{c.c.}$.
Thus the particle trajectories $(x(x_0;t),y(y_0;t))$ satisfy equations

\begin{eqnarray}
\label{particle00}
\frac{d x}{d t} &=u=& Y, \\
\frac{d y}{d t} &=v=& a_0(t)\sin(kx+\phi(t)),
\label{particle01}
\end{eqnarray}
where ${\bf x_0}\equiv(x_0,y_0)$ is the initial particle position
and $a_0(t)$ and $\phi(t)$ are such that $ik A_1\equiv a_0(t)e^{i\phi(t)}$.

\begin{figure}
\centerline{\epsfxsize=6.7cm\epsfysize=6.5cm
\epsfbox{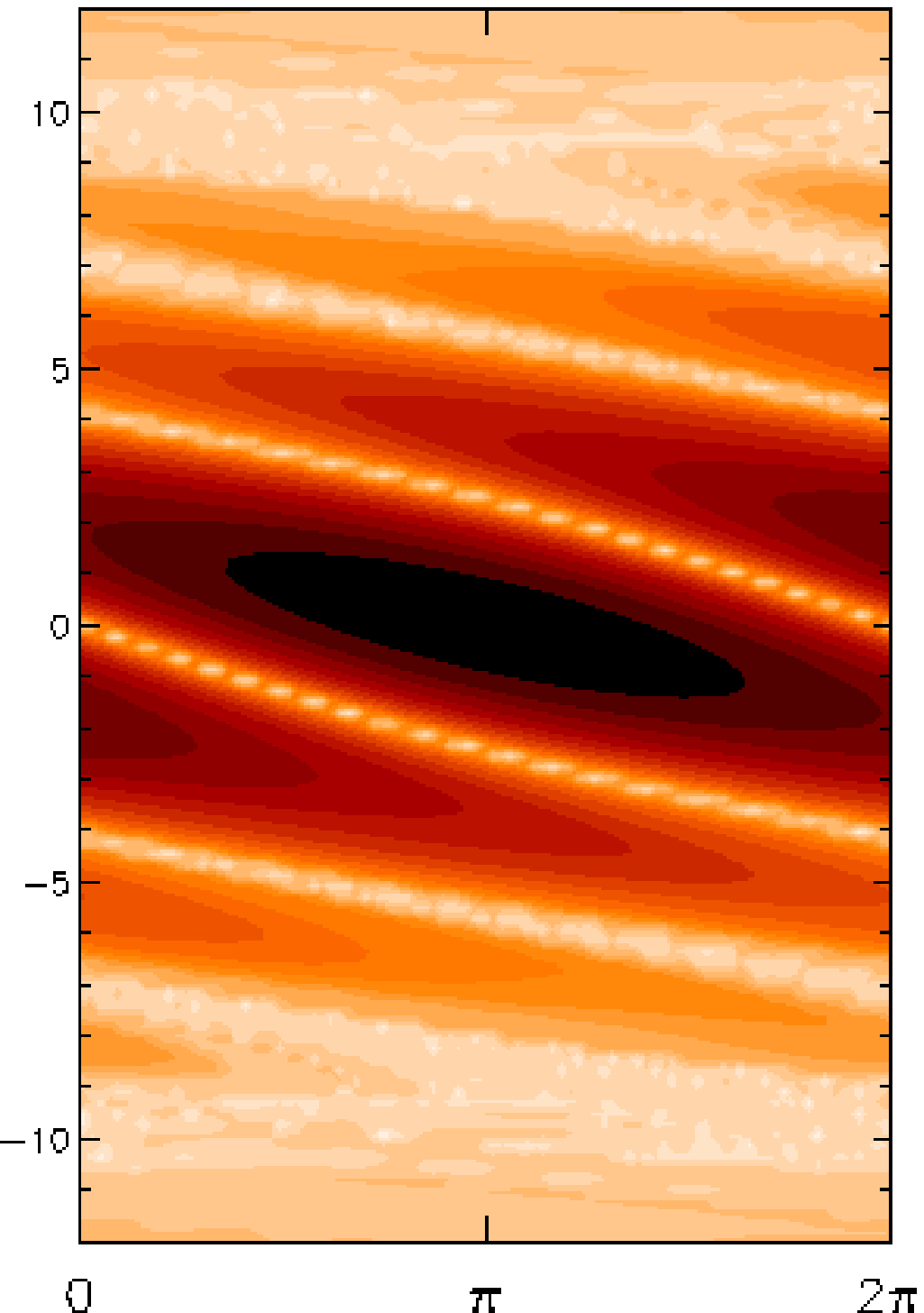}
\hspace{0.3cm}\epsfxsize=6.7cm\epsfysize=6.5cm
\epsfbox{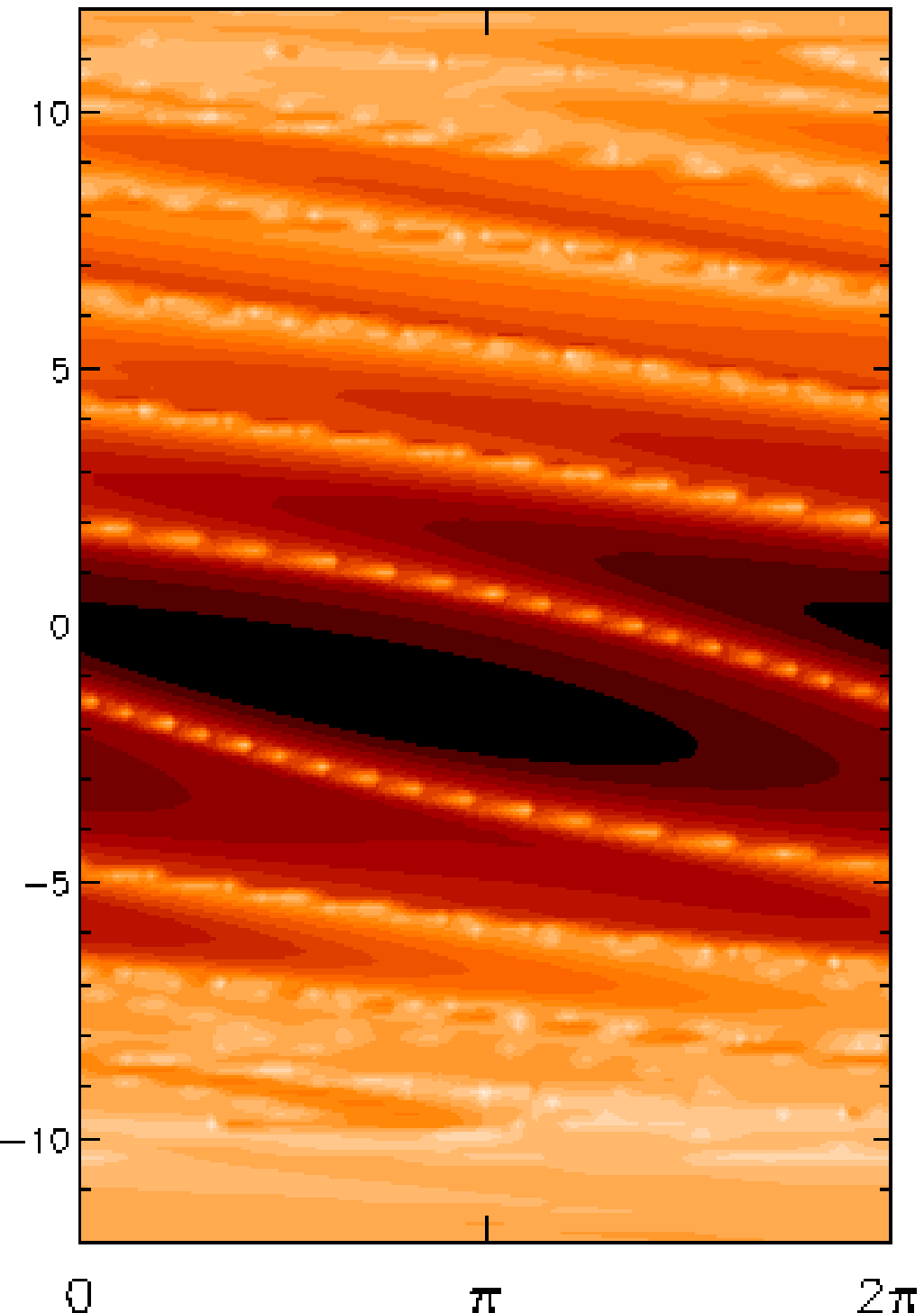}
\hspace{0.3cm}\epsfxsize=1.0cm\epsfysize=6.5cm
\epsfbox{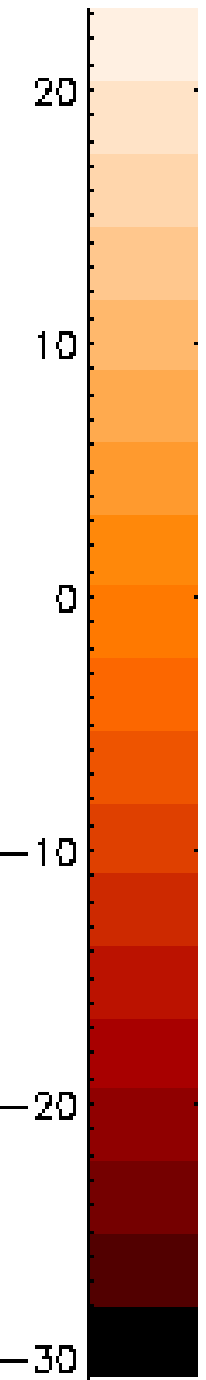}}
%\hspace{1cm}\epsfxsize=4.0cm\epsfysize=5.5cm
%\epsfbox{SRate_pendulum01.eps}}
\caption{
(a):  $\langle \lambda_0 \rangle$ for the critical layer without
coupling to the gravity wave (${\mathcal D}_2=0$ in \citep{Balmforth01}).
(b):  $\langle \lambda_0 \rangle$ for the critical layer coupled to the
gravity wave (${\mathcal D}_2=0.2$).
%(c):  $\langlr \lambda_0 \rangle$ for the pendulum.
}
\label{strainfig}
\end{figure}

If $A_1$ is constant in time, equations
(\ref{particle00})-(\ref{particle01}) reduce to the equations of motion
for a pendulum, only in this case the particle trajectory is not
restricted to a cylindrical surface, and can move from eddy to
eddy. We can calculate the strain rate of such a flow by first linearizing
equations (\ref{particle00})-(\ref{particle01}) for an infinitesimal
separation between two particles $\delta {\bf x}$,

\begin{eqnarray}
\frac{d \delta{\bf x}}{dt} = (\delta {\bf x}\cdot \nabla) {\bf v}
&=&\left( \begin{array}{cc}
                           0  & 1 \\
                           ka_0\cos (kx({\bf x_0};t)+\phi)& 0\end{array}
    \right)\delta {\bf x} \quad .
\label{strainrate}
\end{eqnarray}
$\lambda_0$, which is the negative determinant of the matrix in 
equation (\ref{strainrate}),
is interpreted as the combination of strain and rotation:
$\lambda_0({\bf x_0})=k a_0\cos(kx({\bf x_0};t)+\phi)$. After time-averaging
over the tracer trajectories, $\langle \lambda_0 \rangle$ 
is only a function of  the initial
position of the tracer, and positive $\langle \lambda_0 \rangle$ implies
the possibility for a positive Lyapunov exponent for that initial 
position; negative $\langle \lambda_0 \rangle$
implies that rotation is dominant over strain. We have calculated the
time-averaged $\langle \lambda_0 \rangle$ for each initial position
${\bf x_0}$ in the vorticy for a simple case in
\cite{Balmforth01} (figure \ref{strainfig}(a)).
The corresponding plot for the case of coupling
the shear flow to the gravity surface waves is shown in figure 
\ref{strainfig}(b).
%, together with

\begin{figure}
\centerline{\epsfxsize=6.7cm\epsfysize=6.5cm
\epsfbox{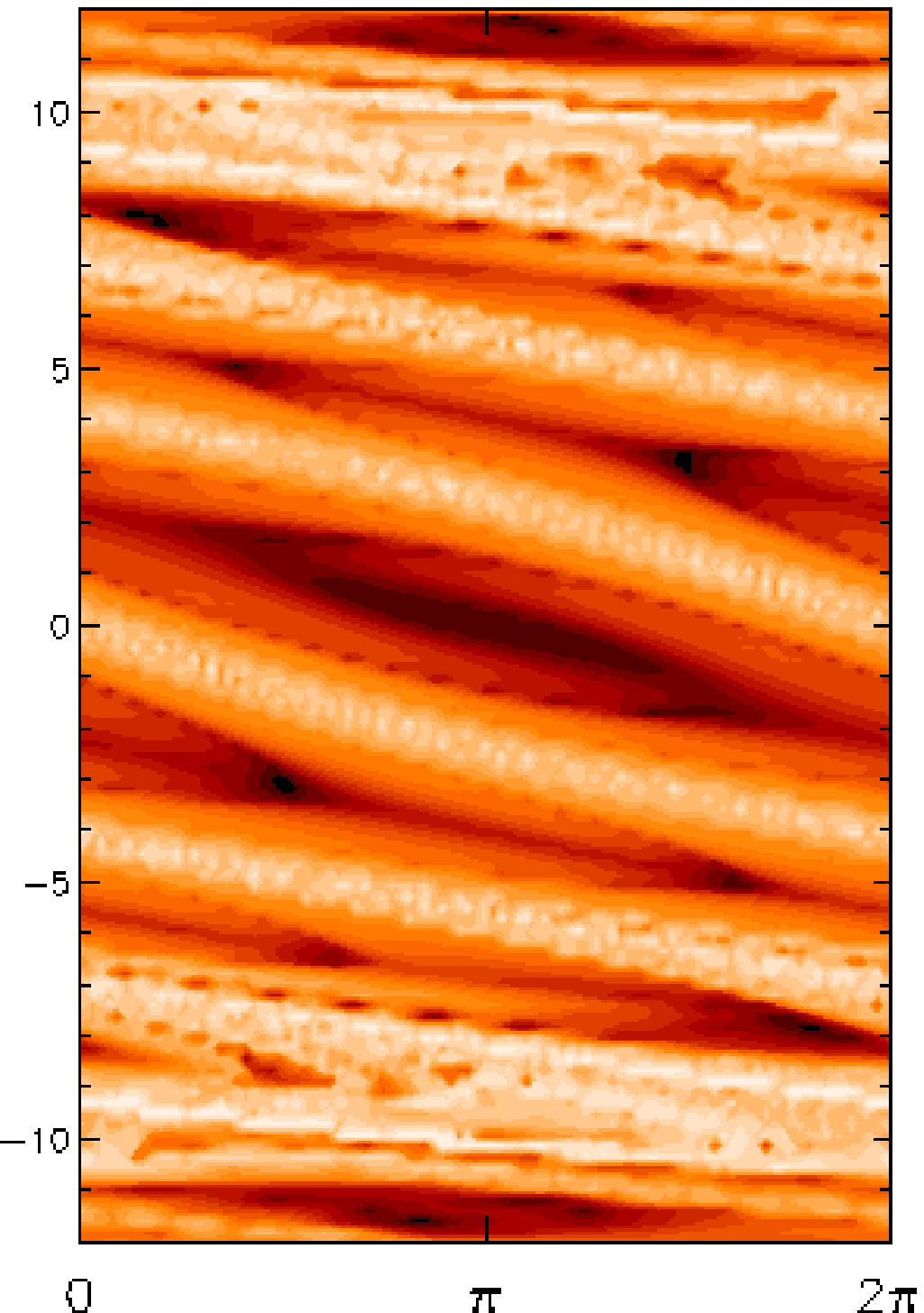}
\hspace{0.3cm}\epsfxsize=6.7cm\epsfysize=6.5cm
\epsfbox{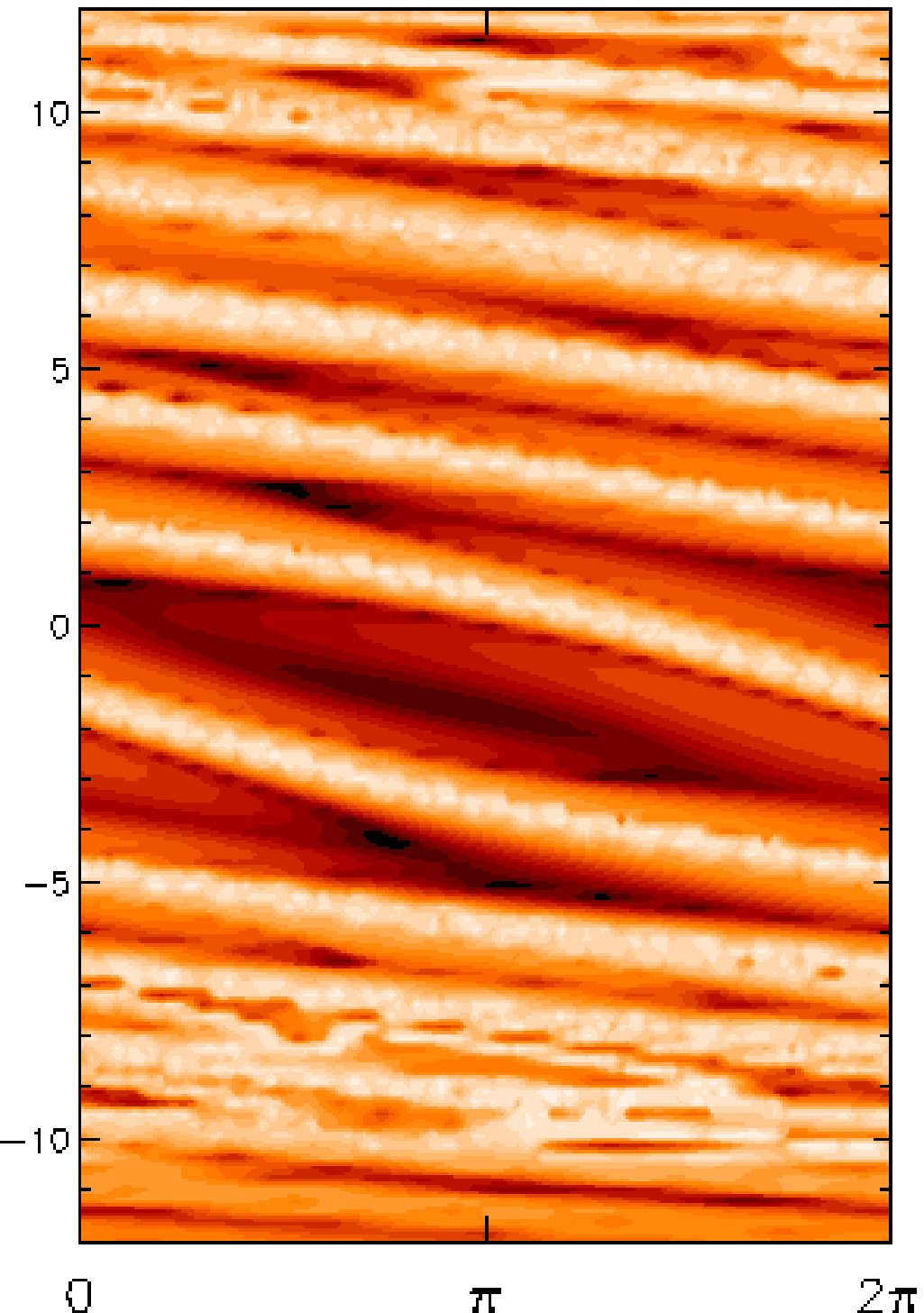}
\hspace{0.3cm}\epsfxsize=1.0cm\epsfysize=6.5cm
\epsfbox{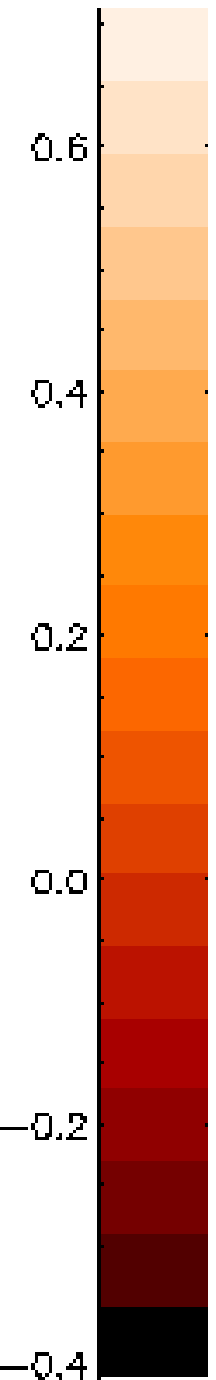}}
\caption{
(a):  Finite time Lyapunov exponent the un-coupled case ${\mathcal D}_2$
in \cite{Balmforth01}.
(b):  Finite time Lyapunov exponent for our case ${\mathcal D}_2=0.2$.
}
\label{Lyapunov}
\end{figure}

The relation between $\langle \lambda_0 \rangle$ and the finite time
Lyapunov exponent may not be straightforward, and can depend sensitively
on the prescribed flow.  In general, some correction to 
$\langle \lambda_0 \rangle$ can
be made so that it is closer to the Lagrangian description
\citep{Boffetta01}.

\begin{eqnarray}
\lambda &=& \langle \lambda_0 \rangle +\langle \lambda_1 \rangle ,\\
\lambda_1 &=& \sqrt{\frac{d \psi_{xy}^2}{dt}
-\frac{d\psi_{xx}}{dt}\frac{d\psi_{yy}}{dt}},
\label{lambda1}
\end{eqnarray}
where $\psi$ is the stream function of the flow. In our case
$\lambda_1=0$, and thus we expect $\langle \lambda_0 \rangle$ to be a
good indicator of the finite time Lyapunov exponent, which we have also
calculated and shown in figure (\ref{Lyapunov}) for the two cases in
figures (\ref{strainfig})a,b.
 
 From figures (\ref{strainfig})-(\ref{Lyapunov}) we conclude that the
coupling of shear flow to the surface wave induces nonlinear evolution in
such a fashion that more chaotic mixing may occur in the latter case; this
conclusion follows from the observation that there appear to be more
stripes of positive Lyapunov exponent in the driven surface wave case.

In situations where the passive tracers are weakly diffusive, the
asymptotic mixing property is determined by  the combination of slow
diffusion and fast advection. If we define $\theta$ as the tracer
concentration, we can write down the equation for the weakly diffusive
tracer in the above flow field, (equations \ref{particle00} and
\ref{particle01}),
\begin{eqnarray}
\label{theta_eq}
\frac{\partial \theta}{\partial t}  + y\partial_x\theta +
(a_0(t)\sin(kx+\phi(t))\partial_y\theta &=& \frac{1}{\Pec} \nabla^2\theta
\,,
\end{eqnarray}
where $\Pec\equiv U_0l/\kappa$ is the Peclet number, with $U_0$ and $l$
the characteristic velocity and length defined in previous section, and
$\kappa$ the molecular diffusivity of the tracer.
%We use a particle method
%to solve equation (\ref{theta_eq}), with $a_0$ and $\phi$ similar to those
%in previous sections, and investigate the time-dependence of particle
%dispersion, which is often indicative of mixing property of the flow.
The particle dispersion (variance), often indicative of mixing property of
parallel flow, is defined as
\begin{eqnarray}
\label{sigma_eq}
\sigma^2 &\equiv& \langle x^2 \rangle - \langle x \rangle^2,
\end{eqnarray}
where $\left<x\right>$ and $\left<x^2\right>$ are, respectively, the 
first and second longitudinal
moments of the concentration field $\theta$
\begin{eqnarray*}
\label{xsq_x}
\langle x \rangle \equiv \int x\theta d^2x,&& \langle x^2 \rangle 
\equiv\int x^2\theta d^2x.
\end{eqnarray*}

If the flow is weak and diffusion is strongly dominant over advection,
particles undergo random walks and their dispersion ($\sigma$) increases
linearly with the square root of time: $\sigma= (2t/\Pec)^{1/2}$. On
the other hand, if the shear flow is strong  and irregular in time, the
particles will be in a super-diffusive regime during which the dispersion
grows faster than that for ballistic transport ($\sigma\sim t$). In cases
where the flow is bounded and time-independent, the super-diffusive regime
eventually gives way to yet another diffusive (Taylor) regime with a
larger effective diffusivity than molecular diffusion.

For a time-dependent velocity field, the super-diffusive regime will
be the asymptotic limit of particle transport, and chaotic mixing may be
found instead.
To examine how the time dependence of the flow affects the
particle mixing in our case, we have integrated equation (\ref{theta_eq})
using a particle method (\citep{Latini01} and references therein) with
$a_0(t)$ and $\phi(t)$ resembling those from solving the amplitude
equations.
We place $10^5$ particles at an initial position close to the separatrix around
the core, solve for their positions according to equation 
(\ref{theta_eq}) for $\Pec = 10^5$,
and record their positions from which we can calculate the particle
dispersion. In our case (equations \ref{particle00} and
\ref{particle01}), ballistic transport is expected (and confirmed
numerically) if there is no time variation in $A_1$.

\begin{figure}
\centerline{\epsfxsize=8.5cm\epsfysize=6.0cm
\epsfbox{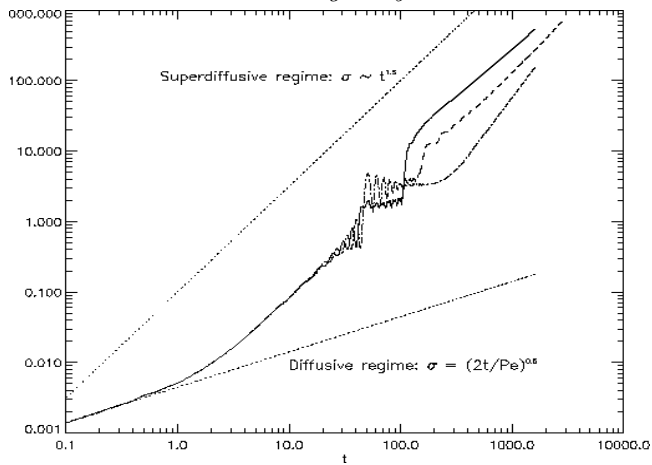}}
\caption{
Particle dispersion for three time-varying amplitudes.
}
\label{particledispersion_fig}
\end{figure}

Figures (\ref{particledispersion_fig}) show the dispersion $\sigma$ versus
time for three time dependences of the amplitude $a_0$. The solid line is
for an exponential growing amplitude which saturates to $a_0\sim 1$ at
$t\sim 100$, and then oscillates periodically with a fluctuation
amplitude of $0.2$ and a period of $6$.  The long dashed line is the
same except the oscillatory period is $65$. The dash-dotted line is
the same as the first two cases except the oscillation is replaced by an
algebraic growth proportional to $t^{2/3}$.  We clearly see that the early
time dependence from simulations is verified as the diffusive regime,
where $\sigma=(2t/\Pec)^{1/2}$. We also observe that the long time
asymptotics for all three cases are at least super-diffusive, with
$\sigma\sim t^{3/2}$ for cases $1$ and $2$, while for case $3$, where the
amplitude grows as $t^{2/3}$, the dispersion is close to $\sigma\sim
t^2$.

These results are consistent with anomalous diffusion found in other
two-dimensional flows \citep{Venkataramani98}. The temporal periodicity
in the flow creates the possibility for KAM regions to coexist with
chaotic regions, and thus the anomalous particle transport.  However, for
the third case, the time variation of the flow is not periodic, and there
does not exist any KAM regions, and thus the particle transport is more
super-diffusive.

\section{Discussion and Conclusion}

In this paper, we have extended our earlier linear analysis of the
resonant interaction between a wind and interface waves to the weakly
nonlinear regime.

Our principal result is the demonstration that the exponential growth of
unstable resonant waves characteristic of the linear regime transitions to
algebraic growth in the weakly nonlinear regime, with the wave amplitude
growing as $t^{2/3}$, similar to the cases without coupling with the
gravity waves. We also find that the algebraic growth is linearly
proportional to the weak viscosity inside the critical layer.
 
Before continuing, we return to examine our assumptions which allowed us to
carry out the weakly nonlinear analysis.  As noted earlier, our analysis is carried
out around the point where the most unstable mode becomes marginally stable
(which corresponds to, for example, the condition ${\mathcal T} G=1/4$ for $r = 0$).
In terms of dimensional parameters, this means that we conduct our analysis for
values of the maximum wind velocity $U$ in the range
\begin{equation}
U_{\rm 1} < U < U_{\rm 2} \,,
\end{equation}
where $U_{\rm 1} $ is found from the criterion that all the perturbations
are neutral in the presence of surface tension for $r=0$:
\begin{eqnarray}
\label{tgcond}
{\mathcal T}G = \frac{\sigma}{\rho_2 U_{\rm 1}^2l}\frac{gl}{U_{\rm 1}^2}&=&\frac{1}{4} \,,
\end{eqnarray}
and $U_{\rm 2} $ is defined as the upper bound on the maximum wind speed for which
only one of the unstable modes has crossed the stability boundary.
This latter condition arises from the assumption of a periodic domain, which leads
to a discrete spectrum, and is crucial in numerical simulations.
%For a periodic domain that is of the same order
%with the wind length scale $l$, $U_{max}$ is a few times larger than $U_{min}$.
%This implies that we need to be very
%careful in applying the conclusions derived here to cases (such as in the laboratory,
%or in the field) for which the periodic domain assumption does not apply.
 
First we note that equation \ref{tgcond} gives a lower 
bound on maximum wind velocity ($U_{\rm 2}$) that is independent of the length scale ($l$) 
of the wind profile. 
Equation (\ref{tgcond}) thus gives a general lower bound on the maximum wind velocity $U$ for
which our analysis applies. For the air-on-water case, we substitute
$\rho_2=1 \gmperc3$, $g=980 \cmpers2$, and $\sigma=72.8 \dynpercm$ into
the above equation, and obtain a lower bound for the wind velocity: $U_{\rm 1} \sim 0.2 \mpers$.
For $U$ below this value in air-on-water system, system is stable \citep{Alexakis01}.
The upper bound on the maximum wind velocity so that our analysis applies 
can be explicitly expressed in terms of size of 
the periodic domain $l_b$, $\sigma$, $g$ and $\rho_2$  
\begin{eqnarray}
\label{tgcond2}
U_{\rm 2}(l_b) &\sim& max\left(\sqrt{\frac{4\pi^2\sigma+g\rho_2 l_b^2}{2\pi\rho_2 l_b}},
                  \sqrt{\frac{16\pi^2\sigma+g\rho_2 l_b^2}{4\pi\rho_2 l_b}}\right),
\end{eqnarray}
where $l_b$ is of the same order as the wind characteristic length: $l_b\sim {\mathcal O}(l)$.
For the air-water case, we find that the upper 
bound $U_{\rm 2}$ to be a few times larger than $U_{\rm 1}$, 
and thus we expect that our results will be a good approximation for winds 
with velocity up to $U_{\rm 2}\sim 1 \mpers$. Winds of this magnitude cause
ripples on water-surfaces but not any wave breaking. 

Within the range of validity of our analysis, we can further examine when the
evolution transitions from  linear to nonlinear regimes, e.g., the transition from
exponential to algebraic growth.  For small density ratio (the air-over-water
case), and for high Reynolds numbers (for which the results from \S 5.3 hold),
this transition happens
when the amplitude is $A_2 \simeq 60$. Re-constructing the dimensional
amplitude, we conclude that the perturbation grows exponentially
until its amplitude reaches $h/l \sim 60 \epsilon^2 \simeq 6\cdot
10^{-5}$, where $l$ is the length scale of the wind (see \S 2; $l$ is of the
same order as the most unstable wavelength) and $\epsilon=r=10^{-3}$ is the
air-over-water density ratio. If, on the other hand, the rescaled Reynolds
number is small enough so that the quasi-steady state approximation (\S5.2)
is appropriate, then the transition will happen when the Haberman
parameter $\lambda$ is close to one ($\lambda \simeq 1$; see
figure \ref{Habberf}). This implies that the transition amplitude
scales as $h/l \sim 1/Re$, which is still very small since we are in the
almost inviscid limit. The key point, however, is that in both cases, the
amplitude at which the transition in behavior occurs is extremely small,
and most likely is too small to be observed in numerical simulations.  Thus, any
linear growth observed in simulations can only result from wind with maximum
velocity much greater than $0.2 \mpers$.  Our results also indicate that the energetic
estimate provided by \cite{Miles99} should only work only when the wind is
much stronger than $0.2 \mpers$.

Furthermore, we note that the linear regime (e.g., exponential growth) would be
observable if the wind velocity is larger than order one. For example, if
$1\gg G^{-1} \gg r$, then the linear growth rate will be larger than order
$r$; this implies a different scaling than $\epsilon=r$. (The
derivation of the new scaling is straightforward, but depends on the how
the free parameter $G$ scales with $\epsilon$ (e.g., $G\sim
\epsilon^{\alpha}$); since this is essentially unknown, we have chosen not
to pursue this calculation further.)\ \ In cases for large $U$, we would
then expect both the exponential and the algebraic growth rates to be
observable in carefully conducted laboratory or numerical experiments.
 
For astrophysical cases, for which the density ratio is not so small
($r\simeq 0.1 \sim 0.5$), it is much easier to capture both scalings
(exponential and algebraic), in either numerical and experimental
work. That is, following the previous arguments, the transition amplitude
between the two scaling regimes is $h/l \sim {\mathcal O}(0.1)$, and can  be
easily varied by changing the magnitude of the wind. Numerical simulations
are therefore obtainable to follow the fully non-linear development of the
waves; and this is what we intend to do in our future work.  Furthermore,
it remains an open question as to whether the above discussion applies to
situations in which the physical domain is not periodic, such as in laboratory
experiments.

Finally, we note that we have obtained an interesting result, namely the
enhanced mixing at the `saddle points' separating successive cat's eyes
within the unstable interface; this enhanced mixing (which we studied by
means of inserting Lagrangian tracers, and observing their evolution) is a
consequence of the dynamics near the `saddle', where mixing between two
adjacent vortices appears to take place.
These islands around the separatrices
are commonly found in non-integrable Hamiltonian systems, and here
they serve as implication of chaotic mixing.
Results from the particle analysis further confirm that chaotic mixing
is a consequence of the temporal behavior of the amplitude associated with the
global background flow.  Even though we have assumed much simpler temporal behavior
for the amplitude in our simulations,
the super-diffusion found in the simpler cases affirms that
chaotic mixing should be expected as a result of
the instability of the shear flow coupled to the gravity surface waves.
This may imply that the entrainment rate of humidity into air could be
enhanced by the coupling of weak wind with suface waves.
 
The calculations presented here will serve as a useful constraint for
future numerical simulations of wind-driven interface instabilities.
Thus, paradoxically, while tests of numerics involving the linear behavior of such
systems can be done, they are challenging to apply; in contrast, it
should be relatively easy to test the numerics in the weakly nonlinear
limit.

%\section{Acknowledgments}
This work was supported in part by the DOE-funded ASCI/Alliances Center
for Thermonuclear Flashes at the University of Chicago.  We thank N.J. Balmforth,
N.R. Lebovitz and S.C. Venkataramani for helpful conversations.  
YNY acknowledges support from NASA and Northwestern University, 
and computation support from Argonne National Labs.

\clearpage

\appendix

\section{Large $G$ behavior}

%                   %   W K B J %

%     LINEAR WIND DRIVEN WAVES FOR THE LARGE G CASE  %

We are interested in cases for which the factor $G$ is large. As already
discussed, such cases not only correspond to the astrophysical limit of
strong surface stratification, but also correspond to cases for which the
growth rate of (linearly) unstable modes is small, i.e., to cases for
which our analysis is actually appropriate.

\begin{figure}[!htb]
\centerline{\epsfysize = 3.in \epsffile{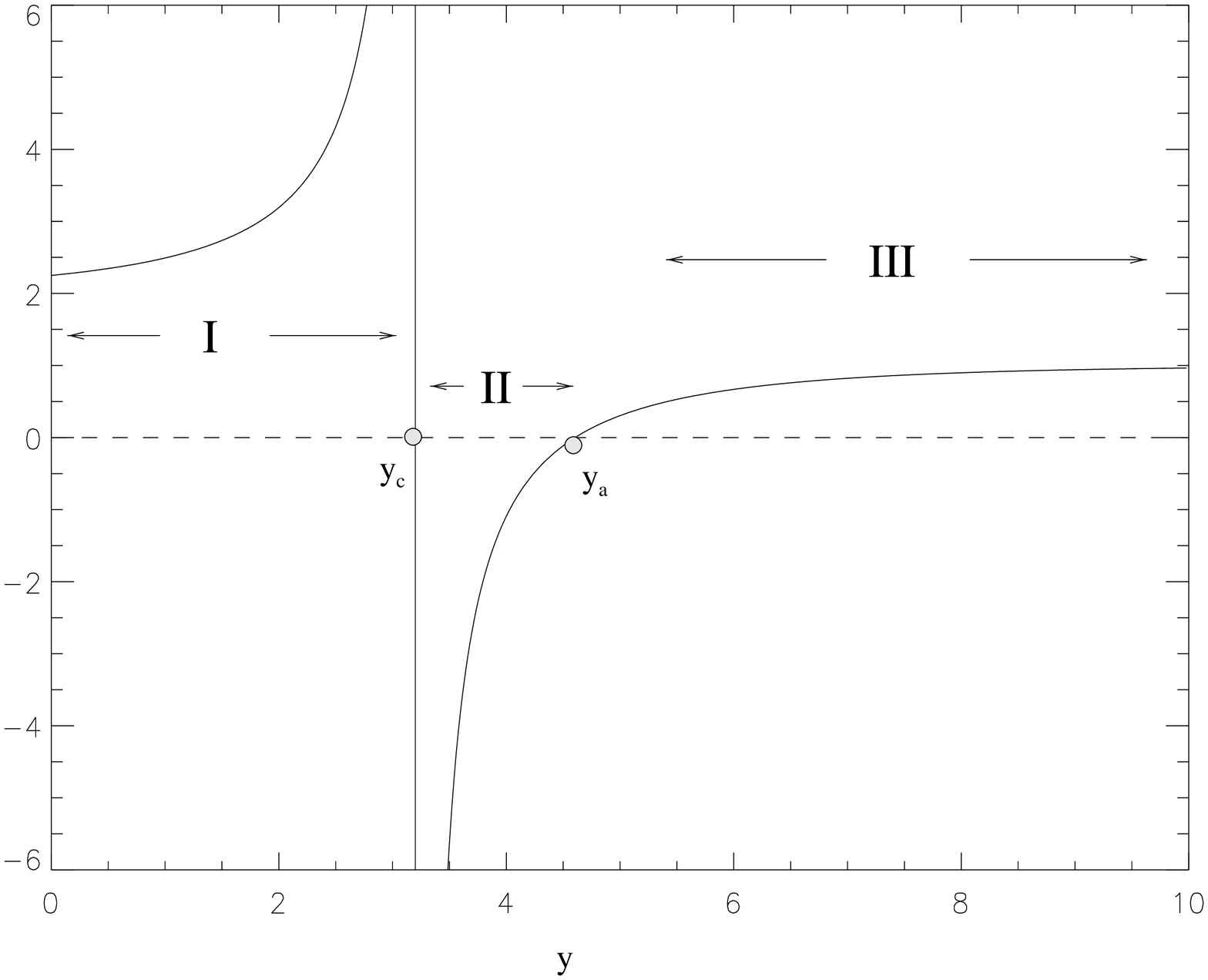}}
\caption{The function F.}
\label{FF}
\end{figure}

In order to proceed, we need to adopt a specific wind profile; in what
follows, we will use a profile of the form $U=U_{max}(1-e^{-y})$. However,
we note that our basic results hold for more general wind profiles. From
equation \ref{BOUND} we know that the unstable  modes will have to be of
the same order as $G$, and therefore $K \gg 1$; this allows us  to
write a WKBJ expansion for the solution of the perturbation stream
function of equation (\ref{linear}) for the wind. The equation we have to
solve for large $K$ is therefore

\begin{equation}
\phi_{,yy}-\left[ K^2+\frac{U_{,yy}}{U-C}\right]\phi=0 \,;
\end{equation}
by rescaling $y \to Ky$, we have
\begin{equation}
\phi_{,yy}-\left[ 1+\frac{1}{K^2}\frac{U_{,y/K,y/K}}{U-C}\right]\phi=0
\end{equation}
or
\begin{equation}
\phi_{,yy}-F(y/K)\phi=0 \,,
\end{equation}
where $F$ is a slowly varying function of its argument, as illustrated in
figure \ref{FF}. The boundary condition at the interface is:
\begin{equation}
KC^2-r\big{[}KC^2\phi_y|_o-CU'|_0\big{]}-G(1-r)=0 \,.
\label{boundary}
\end{equation}
From figure \ref{FF} it is obvious that the WKBJ approximation will break down at
two points: The first one is at $y=y_c$, where the critical layer is
located (and where $F$ becomes singular); the second one is at $y=y_a$,
%(named after Airy)
where $F=0$. For this reason,  we will have to decompose the
$y-$axis into three regions: I. $0<y<y_c$; II. $y_c<y<y_a$; III.
$y_a<y$. The first-order solutions of the WKBJ equations therefore are:
\begin{eqnarray}
\mbox{ (I)} & \quad
\phi=A_1\frac{1}{\sqrt{w}}e^{-\int_0^ywdy'}
       +B_1\frac{1}{\sqrt{w}}e^{+\int_0^ywdy'} \,, \\
\mbox{ (II)} &\quad
\phi=A_2\frac{1}{\sqrt{w}}\sin\left(\int_{y_c}^ywdy'-\pi/4\right)
+B_2\frac{1}{\sqrt{w}}\cos\left(\int_{y_c}^ywdy'-\pi/4\right) \,, \\
\mbox{ (III)} &\quad
\phi=A_3\frac{1}{\sqrt{w}}e^{-\int_{y_a}^ywdy'} \,,
\end{eqnarray}
where
\[
w(y)=\sqrt{\left| 1+\frac{1}{K^2} \frac{U_{,y/K,y/Ki}}{U-C} \right|}
=\sqrt{\left| 1+\frac{1}{K^2}\frac{e^{-y/K}}{1-C-e^{-y/K}} \right|} \,,
\]
and the $-\pi/4$ factor appearing in the solution for Region II is
inserted for convenience, to be exploited shortly.
The coefficients $A_1,B_1,A_2,B_2,A_3$ are connected through the
solutions at the points where the WKBJ approximation breaks down, and can
be obtained using matched asymptotics. Thus, close to $y=y_a$ it is
well-known that the solution is an Airy function; the consequent
connection formulae are given below \citep{Olver}. For $y < y_a$,
\begin{eqnarray*}
\phi&=&\frac{2A_3}{\sqrt{w}}\sin\left[-\int^y_{y_a}wdy'+\pi/4\right]
   \,, \\
%&=&\frac{2A_3}{\sqrt{w}}\sin\left[\int^{y_a}_{y_c}wdy'
%-\int^y_{y_c}wdy'+\pi/4 \right] \,, \\
%&=&\frac{2A_3}{\sqrt{w}}\sin\left[I_1-\int^y_{y_c}wdy'+\pi/4 \right] \,,
%\\
&=&\frac{2A_3}{\sqrt{w}}\left\{
\sin\left[I_1\right]\cos\left[\int^y_{y_c}wdy'-\pi/4 \right]-
\cos\left[I_1\right]\sin\left[\int^y_{y_c}wdy'-\pi/4 \right]
\right\} \,,
\end{eqnarray*}
and therefore
\begin{eqnarray}
B_2=2A_3\sin\left[I_1\right],&&
A_2=-2A_3\cos\left[I_1\right] \,,
\end{eqnarray}
where
$I_1=\int^{y_a}_{y_c}wdy'$.
The solutions near the critical point $y=y_c$ satisfy the equation
\begin{equation}
\phi_{,yy}-\frac{U''_c}{KU'_cy}\phi=0
\label{bes}
\end{equation}
The solutions of the above equations are given in terms of
$z=-(y-y_c)U_c''/U_c'K >0$ by
\begin{eqnarray}
f_1(z)=\sqrt{z}J_1\left(2\sqrt{z}\right), &&
f_2(z)=\sqrt{z}N_1\left(2\sqrt{z}\right),
\end{eqnarray}
where $J_1,N_1$ are, respectively, the first Bessel and Neumann (Bessel of
the second kind) functions. The first terms in the asymptotic expansion
for $y\to +\infty$ are
\begin{eqnarray}
f_1(z) \simeq
\frac{\sqrt[4]{z}}{\sqrt{\pi}}\sin\left(2z^{1/2}-\pi/4\right), && 
f_2(z) \simeq
-\frac{\sqrt[4]{z}}{\sqrt{\pi}}\cos\left(2z^{1/2}-\pi/4\right).
\end{eqnarray}
Matching with the outer solution, we have
\[
\phi=\sqrt{\pi}A_2f_1(z)-\sqrt{\pi}B_2f_2(z) \,.
\]
The asymptotic expansion for $z\to 0^+$ is
\begin{eqnarray}
f_1 &\simeq & z+\dots \\
\pi f_2 &\simeq & -1+z\ln|z|+\dots-(1-2\gamma)z+\dots \,,
\end{eqnarray}
where $\gamma$ is the Euler-Masceroni constant. Thus, we can identify
$f_1$ with the regular Frobenius solution $\phi_a$,
and $f_2$ with the singular Frobenius solution $\phi_b$.
Here we assumed that $C_i$ is much smaller that $G^{-1}$ (this is
something that will be justified a posteriori).

For $y<y_c$ the solutions of equation (\ref{bes}) can be obtained by
making the transformation $z \to e^{-i\pi}z$ which is equivalent to
taking the contour below the critical layer. By doing this the first
Bessel function transforms to the first modified Bessel function that is
growing exponentially and is real while the second one transforms to a
linear combination of an exponentially growing and an exponentially
decreasing modified Bessel function, and due to the presence of the
logarithm it is going to have an imaginary part. Their asymptotic
expansion for $y\to \infty$ is
\begin{eqnarray}
f_1 \simeq 
-\frac{\sqrt[4]{z}}{2\sqrt{\pi}}e^{+2\sqrt{z}},&& 
f_2 \simeq
\frac{\sqrt[4]{z}}{2\sqrt{\pi}}\left[ie^{2\sqrt{z}}-e^{-2\sqrt{z}}\right],
\end{eqnarray}
with  $z=(y-y_c)U_c''/U_c'K >0$. The inner solution  for negative
large $z$ can be then written as

\begin{eqnarray}
\phi_{in}&=&\sqrt{\pi} \left(A_2f_1-B_2f_2 \right) \,, \\
%&=&\frac{\sqrt[4]{z}}{2}\left[
%-A_2e^{2\sqrt{z}}-B_2\left(ie^{2\sqrt{z}}-e^{-2\sqrt{z}}
%\right)\right] \,, \\
&=-&\frac{\sqrt[4]{z}}{2}\left[
\left(A_2+iB_2 \right)e^{2\sqrt{z}}-B_2e^{-2\sqrt{z}}\right] \,.
\end{eqnarray}
Matching with the outer solution then gives

\begin{eqnarray}
\phi &=& \frac{1}{2\sqrt{w}}\left[
-(A_2+iB_2)e^{-\int_{y_c}^ywdy'}+
B_2e^{+\int_{y_c}^ywdy'}\right] 
%&=&
%\frac{1}{2\sqrt{w}}\left[
%-(A_2+iB_2)e^{+\int^{y_c}_0wdy'}e^{-\int_{0}^ywdy'}+
%B_2e^{-\int^{y_c}_0wdy'}e^{+\int_{0}^ywdy'}
%\right] \,,
\end{eqnarray}
or
\begin{equation}
\phi=-\frac{A_2+iB_2}{2}e^{I_2}\frac{1}{\sqrt{w}}e^{-\int_{0}^ywdy'}+
\frac{B_2}{2}e^{-I_2}\frac{1}{\sqrt{w}}e^{\int_{0}^ywdy'} \,,
\end{equation}
where
$I_2=\int^{y_c}_0wdy'$.
Gathering all the terms then gives
\begin{eqnarray}
A_1=A_3\bigg{(}\cos\left[I_1\right]
            -i\sin\left[I_1\right]\bigg{)} e^{+I_2}, &&
B_1=A_3\sin\left[I_1\right]e^{-I_2},
\end{eqnarray}
and
\begin{equation}
\phi=A_3\left[
\sin(I_1)e^{-I_2}e^{+\int_{0}^ywdy'}+
\cos(I_1)e^{+I_2}e^{-\int_{0}^ywdy'}-
i\sin(I_1)e^{+I_2}e^{+\int_{0}^ywdy'}
\right].
\end{equation}
The values of $\phi$ and $\phi_{,y}$ at zero are therefore
\begin{eqnarray}
\phi|_0&=&A_3\left[\sin(I_1)e^{-I_2}+\cos(I_1)e^{+I_2}-
i\sin(I_1)e^{+I_2} \right],\\
\phi_{,y}|_0&=&A_3\left[\sin(I_1)e^{-I_2}-\cos(I_1)e^{+I_2}+
i\sin(I_1)e^{+I_2} \right],
\end{eqnarray}
where we have kept terms only to first order in $K$. For the given wind
profile ($1-e^{-x}$), we have:
\begin{eqnarray*}
y_c=-K\ln(1-C), &&
y_a=-[K\ln(1-C)-K\ln(1+1/K^2),
\end{eqnarray*}
\begin{equation}
I_1 \simeq \frac{\pi}{2K}+{\mathcal O}(1/K^3) \,,
\end{equation}
\begin{equation}
I_2 \simeq y_c+O(1/K^2)=-K\ln(1-C)+{\mathcal O}(1/K) \,,
\end{equation}

\begin{eqnarray*}
\sin(I_1)\simeq \frac{\pi}{2K}, \,\,&\quad \cos(I_1) \simeq 1,&\,\,
e^{-I_2}=(1-C)^K.
\end{eqnarray*}

Normalizing so that $\phi|_0=1$, we then obtain
\begin{equation}
\phi_{,y}=
\frac{\sin(I_1)e^{-I_2}-\cos(I_1)e^{+I_2}+i\sin(I_1)e^{+I_2}}
{\sin(I_1)e^{-I_2}+\cos(I_1)e^{+I_2}-i\sin(I_1)e^{+I_2} } \,,
\end{equation}
or
\begin{equation}
\phi_{,y}=-1+
\frac{2\sin(I_1)e^{-I_2}}
{\sin(I_1)e^{-I_2}+\cos(I_1)e^{+I_2}-i\sin(I_1)e^{+I_2} } \,.
\label{WKBFy}
\end{equation}
The second term in equation (\ref{WKBFy}) is exponentially small
when compared to $1$ since $I_2 \sim K$; neglecting this term when
appropriate then allows $\phi_{,y}$ to be written as

\begin{equation}
\phi_{,y}=-1+ 2i\sin^2(I_1)e^{-2I_2} \,,
\end{equation}
where we have kept only the first term in the expansion of the real and
imaginary parts. Plugging in this value of $\phi_{,y}$ in
equation (\ref{boundary}), we obtain, to zeroth order,
\begin{equation}
C_0=\sqrt{\frac{1-r}{1+r}\frac{G}{K}}=\sqrt{{\mathcal{A}}_tG/K} \,,
\end{equation}
which corresponds to the gravity wave in the absence of a wind;
${\mathcal{A}}_t$ is the Atwood number. For our purposes, this is as far
as we need to go in analyzing the real part of $C$.

We next turn to analyzing the imaginary part of $C$. To obtain the first
order in Im$\{C\}=C_i \ll C_0$ we have:
\begin{eqnarray}
2K(1-r)C_0C_i-rKC_0^2(1+\phi_{,y})=0,&& 
C_i=\frac{1}{2}\frac{r}{1-r}C_0(1+\phi_{,y}),
\end{eqnarray}
so that

\begin{equation}
C_i
=\frac{rC_0}{1-r} \sin^2(I_1)e^{-2I_2},
\label{Idep}
\end{equation}
\begin{equation}
\mbox{Im}\left\{C_1\right\}
=\frac{r\pi^2}{4(1-r)}\frac{1}{{\mathcal{A}}_t^2G^2}
\left(\frac{{\mathcal{A}}_tG}{K}\right)^{5/2}
\left[\left(1-\frac{1}{\sqrt{K/({\mathcal{A}}_tG)}}\right)^
{(K/{\mathcal{A}}_tG)}
\right]^{2{\mathcal{A}}_tG}.
\end{equation}
We note that $(1-1/\sqrt{x})^{2x}$ is a bounded function smaller than
one, and therefore $C_i$ has a negative exponential dependence on
$G$. We note further that this exponential dependence should be
independent of the wind profile, and in a more general case --- for which
$U(y)$ is the wind profile and $U^{-1}(c)$ is its inverse --- the growth
rate will be proportional to $C_i \sim \exp [-2KU^{-1}(c)]$; this can
be re-written as $C_i \sim f(c(K))^{{\mathcal{A}}_tG}$, with $f(c) \equiv
\exp [-2U^{-1}(C)/C]$ a bounded function and $C=C_0$. The interpretation
of equation (\ref{Idep}) is straightforward: it simply states that the
growth rate is proportional to the negative exponential of the height of
the critical layer, as measured in units of the wavelength.

\bibliography{wdgw}

\end{document}